\documentclass[12pt]{article}
\usepackage{amsmath,amssymb,mathrsfs,alpheqn,epsfig}
\usepackage[a4paper]{geometry}
\newcommand{\rf}[1]{(\ref{#1})}
\newcommand{\tr}{\operatorname{tr}}

\newcommand{\A}{{\mathcal{A}}}
\newcommand{\bs}{\boldsymbol}
\newcommand{\CC}{{\mathbb{C}\,}}

\newcommand{\D}{{\mathcal{D}}}
\newcommand{\DF}{{\mathcal{D}}_{{\mathrm{F}}}}
\newcommand{\DS}{{\mathcal{D}}_{{\mathrm{S}}}}
\newcommand{\DT}{{\mathcal{D}}_{{\mathrm{T}}}}
\newcommand{\DeF}{\Delta_{{\mathrm{F}}}}
\newcommand{\Din}{\Delta_{{\mathrm{in}}}}
\newcommand{\DeS}{\Delta_{{\mathrm{S}}}}
\newcommand{\DeT}{\Delta_{{\mathrm{T}}}}
\newcommand{\F}{{\mathcal{F}}}
\renewcommand{\H}{{\mathscr{H}}}
\newcommand{\I}{{\mathcal{I}}}
\newcommand{\J}{{\mathcal{J}}}
\newcommand{\K}{{\mathcal{K}}}
\renewcommand{\L}{{\mathcal{L}}}
\newcommand{\Lin}{L_{{\mathrm{in}}}}
\newcommand{\M}{{\mathcal{M}}}

\newcommand{\oPsi}{\overset{{\bs\circ}}{\Psi}}
\newcommand{\p}[1]{{{\mathscr{D}}}^{(#1)}}
\renewcommand{\P}{{\mathscr{P}}}

\newcommand{\paraN}{\,{}^{(||)}N}
\newcommand{\paraQ}{\,{}^{(||)}Q}
\newcommand{\perpN}{\,{}^{(\perp)}N}
\newcommand{\perpQ}{\,{}^{(\perp)}Q}
\newcommand{\paraB}{\,{}^{(||)}B}
\newcommand{\psiI}{\psi_{{\mathrm{I}}}}
\newcommand{\psiII}{\psi_{{\mathrm{II}}}}

\newcommand{\T}{{\mathcal{T}}}
\newcommand{\xin}{x_{{\mathrm{in}}}}
\newcommand{\Y}{{{\mathcal{Y}}}}

\numberwithin{equation}{section}
\renewcommand{\thesection}{\Roman{section}}

\newcommand{\B}{{\mathcal{B}}}
\newcommand{\Bcirc}{{\stackrel{\circ}{\mathcal{B}}}}
\newcommand{\Bdt}{{\stackrel{{\stackrel{\sim}{\sim}}}{\mathcal{B}}}}

\newcommand{\C}{{\mathcal{C}}}
\newcommand{\Ccirc}{{\stackrel{\circ}{\mathcal{C}}}}

\newcommand{\epsilonvec}{{\bs{\epsilon}}}
\newcommand{\GO}{{\mathfrak{so}}}
\newcommand{\E}{{\mathcal{E}}}
\newcommand{\f}{{\mathrm{f}}}
\newcommand{\ff}{{\mathfrak{f}}}

\renewcommand{\H}{{\mathcal{H}}}

\renewcommand{\P}{{\mathcal{P}}}

\renewcommand{\S}{\mathcal{S}}

\renewcommand{\S}{{\mathcal{S}}}

\newcommand{\thetavec}{{\bs{\theta}}}

\newcommand{\Z}{{\mathcal{Z}}}

\begin{document}
\setcounter{page}{0}
\title{\bf Geometry and Topology\\ of\\ Relativistic Two-Particle Quantum Mixtures}
\author{S.\ Hunzinger, M. Mattes, and M.\ Sorg\\[2cm] II.\ Institut f\"ur Theoretische Physik der
Universit\"at Stuttgart\\ Pfaffenwaldring 57 \\ D 70550 Stuttgart, Germany}
\maketitle
\begin{abstract}
\indent
Within the framework of Relativistic Schr\"odinger Theory (an alternative form of quantum mechanics for relativistic many-particle systems) it is shown that a general N-particle system must occur in one of two forms: either as a ``positive'' or as a ``negative'' mixture, in analogy to the fermion-boson dichotomy of matter in the conventional theory. The pure states represent a limiting case between the two types of mixtures which themselves are considered as the RST counterparts of the entangled (fermionic or bosonic) states of the conventional quantum theory. Both kinds of mixtures are kept separated from dynamical as well as from topological reasons. The 2-particle configurations ($N=2$) are studied in great detail with respect to their geometric and topological properties which are described in terms of the Euler class of an appropriate bundle connection. If the underlying space-time manifold (as the base space of the fibre bundles applied) is parallelisable, the 2-particle configurations can be thought to be generated geometrically by an appropriate ($2+2$) splitting of the local tangent space.
\end{abstract}
\newpage
\section{Introduction and Survey}

Surely one of the most striking features of matter refers to the fact that one encounters two distinct types of particles: fermions and bosons. As safely as this fact is established from both the conceptual and experimental point of view (for a review see ref.~\cite{Su90}), it seems unclear for most people what is the precise theoretical origin of this matter dichotomy and why there do not exist more than two kinds of material particles. In view of the lack of a right understanding of the reason why there are only two particle species, some people hold that one must first ``\emph{understand why nature makes such a particular choice from among a host of other possibilities}''~\cite{OhKa82}. Consequently for that purpose, they believe, one should ``\emph{pursue every theoretical possibility we may conceive at present}''~\cite{OhKa82} and these endeavours should refer to the rather general framework of paraquantization. Other people do not trouble themselves about the true origin of the fermion-boson dichotomy but take this simply as an experimentally based fact, and consequently they are content with rephrasing this fact in mathematical terms (i.e.\ the postulates of symmetrization for bosons and anti-symmetrization for fermions). In this sense, e.g., Dirac simply states that ``\emph{other more complicated kinds of symmetry are possible mathematically, but do not apply to any known particles}'' (ref.~\cite{Di98},~p.~211). 

One could go even one step further by resorting to the well-known \emph{anthropic principle}~\cite{Ha88} which may be evoked whenever a well-established observation cannot be based upon a sound theoretical foundation. Such a situation typically occurs in cosmology~\cite{KoTu90,ZhiXi89} but the present fermion-boson dichotomy could also evoke that principle, e.g. by asking: ``{\it Did God - for lack of a better word - build a series of failed worlds which spottered and died, or exploded and disintegrated, before discovering the stabilizing effect of anti-commutation relations for half-integral spin fields?}''~(ref.~\cite{DuSu98},~p.~4). However it rather seems reasonable to assume that the fermion-boson dichotomy ``{\it could conceivably be an essential ingredient of a more fundamental view of the world~\ldots This could be the case in fundamental string theories or their successors~\ldots}''~\cite{DuSu98}.

In any case, such an unclarified situation with the matter dichotomy should provide sufficient motivation for studying this problem also within the context of non-standard quantum theories. In this sense the present paper is devoted to the study of the matter dichotomy within the framework of Relativistic Schr\"odinger Theory (RST), an alternative form of quantum theory being based upon fluid-dynamic concepts rather than probabilistic ones~\cite{So97} -\cite{MaRuSo}. Indeed it has turned out that a matter dichotomy emerges in this theory in the form of positive and negative mixtures~\cite{RuSo01,RuSoIJTP} resembling very much the symmetrized (bosonic) and anti-symmetrized (fermionic) states of the conventional quantum theory: the positive mixtures imply a certain kind of \emph{fusion} of two particles into one charge density, and the negative mixtures describe the recession of the two charge densities from one another such that the density of one particle is zero in that region of space-time where the density of the other particle is non-zero (charge separation~\cite{RuSoIJTP}). Consequently both particles in a negative-mixture configuration are forbidden to occupy the same space-time points which evidently yields a ``stabilizing effect'' for composite matter coming about in the conventional quantum theory through the anti-symmetrization postulate (or anti-commutation postulate for operators, resp.).

However before plunging deeply into the physical implications of this RST matter dichotomy, one first would like to understand better both the  differences and the common features of both types of RST mixtures from a purely \emph{mathematical} point of view.  In this sense, the present paper presents a detailed study of the geometric and topological properties of both types of mixture configurations, deferring the discussion of the physical implications to a separate treatment. For the purpose of a survey of the corresponding results, it is very instructive to first recall some of the basic (probabilistic) concepts of the conventional quantum theory and oppose these to the analogous kinematical elements of the (fluid-dynamic) RST. Since, on the conventional side, there exists no successful \emph{relativistic} quantum mechanics for many-particle systems (in contrast to the very successful quantum field theory), we have to restrict ourselves here to a comparison of the {\it  non-relativistic} conventional theory with RST (as a fully relativistic theory). Furthermore we are satisfied in the present paper with a treatment of 2-particle  systems (neglecting also the particle spin) which may be sufficient to display already the essential many-particle effects. (For including also the spin into RST see, e.g.~ref.~\cite{MaSo99}).

\subsection{Conventional Quantum Theory}

Since in the conventional theory the 2-particle Hilbert space is the tensor product of the two 1-particle spaces, a \emph{disentangled} 2-particle state~$\Psi(1,2)$ is simply the product of two 1-particles states~$\psiI$ and~$\psiII$, i.e.
\begin{equation}
  \label{eq:prod}
  \Psi(1,2) = \psiI(1)\otimes\psiII(2) \; .
\end{equation}
In contrast to this product construction, a disentangled 2-particle state~$\Psi(x)$ in RST is the direct sum of the two 1-particle states
\begin{equation}
  \label{eq:sum}
  \Psi(x) = \psiI(x)\oplus\psiII(x)\ ,
\end{equation}
or reformulated in mathematical terms: the 2-particle vector bundle (with the RST 2-particle wave function~$\Psi(x)$ as bundle section) is the Whitney sum of both 1-particle bundles. If some interaction between the particles is switched-on (e.g.\ the electromagnetic interactions, especially the static Coulomb interaction) the conventional state~\rf{eq:prod} cannot retain its simple product form~\rf{eq:prod} but develops into a general 2-particle state~$\Psi_\pm(1,2)$ obeying the symmetrization postulate
\begin{equation}
  \label{eq:syp}
  \Psi_\pm(1,2) = \pm\Psi_\pm(2,1)\ .
\end{equation}
In contradiction to certain claims in the older literature, the modern textbooks clearly state that \emph{``this postulate} (expressing the matter dichotomy in the conventional theory) {\it cannot be deduced from other principles of quantum mechanics''} (ref.~\cite{Ba98},~p.~475).

If the interparticle interactions are switched-off again, the generally entangled 2-particle state~$\Psi_\pm$~\rf{eq:syp} does not return to the original disentangled form~\rf{eq:prod} but develops to a simply entangled state~$\oPsi$ which is build up again by two 1-particle states but still obeys the symmetrization postulate~\rf{eq:syp}
\begin{equation}
  \label{eq:syp2}
  \oPsi_\pm = \frac{1}{\sqrt{2}}\left(\psiI(1)\otimes\psiII(2)\pm
    \psiII(1)\otimes\psiI(2)\right)\ .
\end{equation}
This form of the wave function persists even when the particles are localized at large spatial separation and thus gives rise to the well-known Einstein-Podolsky-Rosen paradoxes~\cite{Se90}.

\subsection{Relativistic Schr\"odinger Theory}

According to the Whitney sum construction~\rf{eq:sum} (in place of the conventional product arrangement~\rf{eq:prod}), the concept of {\it entanglement} is defined in RST in a somewhat different way, namely by describing matter by means of a (Hermitian) intensity matrix~$\I$ in place of a wave function $\Psi$. The disentangled RST states~$\Psi$~\rf{eq:sum} can also be written in form of an intensity matrix~$\I$ but here~$\I$ adopts a special form, namely the tensor product of~$\Psi$ and its Hermitian conjugate~$\bar{\Psi}$, see equation~\rf{eq:IPsi} below. A disentangled RST state~$\Psi$~\rf{eq:sum} obeys the Relativistic Schr\"odinger Equation~\rf{eq:RSE} and an entangled RST configuration~$\I$ (a ``mixture'') satisfies the Relativistic von Neumann equation~\rf{eq:calI}. The Hamiltonian~$\H_\mu$, governing both these field equations~\rf{eq:RSE} and (\ref{eq:calI}), is itself a dynamical object of the theory and has to be determined from its field equations~\rf{eq:IntCond} and~\rf{eq:ConEq}. All these RST prerequisites, being indispensable for the subsequent mathematical elaborations, have been collected into {\bf Section~II}.

However, the crucial point with the matter dichotomy is now that it arises in RST as a direct consequence of the RST dynamics itself and not as an additional kinematical postulate as in the conventional theory~\rf{eq:syp}. This is proven in {\bf Section~III}  for an arbitrary number~$N$ of (scalar) particles. The RST dichotomy consists in the fact that the determinant of the~$N\times N$ intensity matrix~$\I$ is always of definite sign for any mixture type, see the corresponding discussion of the equation~\rf{eq:DNdetIN} below, together with fig.~1. Thus, in RST there occur exactly two forms of matter, namely the positive and negative mixtures; this is in contrast to the conventional theory which has its problems with the \emph{theoretical} exclusion of the higher-order multiplets of the symmetric group of identical particle permutations \cite{OhKa82}, leaving  over only the symmetric and anti-symmetric states as its lowest-order representations.

After the matter dichotomy has thus been safely established for a general N-particle system, we restrict ourselves for the remainder of the paper to the \emph{2-particle systems (N=2)}. Therefore in {\bf Section IV} all relevant RST mathematics is specialized down to the 2-particle case which requires as the typical fibre of the vector bundle the two-dimensional complex space~$\CC^2$ (and its tensor-product associates). The interesting point with the RST 2-particle kinematics is here that one can reparametrize the whole field system by single-particle variables and by exchange fields. This is again in contrast to the conventional theory which (for an entangled state) does attribute one single 2-particle wave function~$\Psi_\pm(1,2)$~\rf{eq:syp}-\rf{eq:syp2} to both particles but does not attribute a 1-particle wave function to either particle.

As a consequence of this kinematic peculiarity, one can clearly identify in RST the exchange forces (different from the gauge forces) which are responsible for the fusion effects of the positive mixtures ($\to$~bosonic states) and separation effects of the negative mixtures ($\to$~fermionic states). Indeed the mixture variable~$(\lambda)$ enters the single-particle RST dynamics~\rf{e4:44} together with the exchange fields~$(S_\mu,T_\mu,F_\mu,G_\mu)$  in form of a potential term so that one arrives at the disentangled 2-particle situation (i.e. pure RST state~$\Psi(x)$~\rf{eq:sum} analogous to the conventional~$\Psi(1,2)$~\rf{eq:prod}) when one puts the exchange fields to zero. Such a disentangled RST situation can then be described by two ordinary Klein-Gordon equations~\rf{e4:46}, namely one equation for either particle. This RST picture of how the exchange forces produce the fusion and separation effects has no counterpart in the conventional theory, where it is not possible to trace back the Pauli exclusion principle to the action of some kind of force preventing, say, a fermion from occupying the same quantum state as an other identical fermion (see the discussion of this problem in ref.~\cite{Pol}).

But with a convenient parametrization of the 2-particle systems being now at hand, one can study the geometric and topological differences of both kinds of mixture configurations ({\bf Section~V}). Here, the point of departure is the observation that the dynamical equations (of Section~IV) ensure the existence of a closed 2-form~${\mathbf{E}}\ ({\mathbf{dE}}\equiv 0)$ which is composed of the mixture variable~$\lambda$ and the exchange fields~$\mathbf{G}$ and~$\mathbf{F}$, see equation~\rf{e5:1} below. It is just the process of revealing the origin of the closedness of~$\mathbf{E}$ which yields a detailed demonstration of the mathematical peculiarities of the mixtures:

First it can be shown that~$\mathbf{E}$ is the Euler class of an appropriately constructed fibre bundle over space-time (density bundle) so that the closedness relation for~$\mathbf{E}$~\rf{e5:3} appears as nothing else than the Bianchi identity for the bundle curvature. Next it is proven that the winding numbers of the Euler class~$\mathbf{E}$ are always zero and thus~$\mathbf{E}$ is an {\it exact} 2-form over space-time (\textbf{E}=\textbf{d\,e}, see equation~\rf{e5:5}). The origin of the exactness of~$\mathbf{E}$ lies in the fact that~$\mathbf{E}$ is the pullback of the surface 2-cell of the one-parted hyperboloid~$H_{(-)}$ for the negative mixtures and of the two-parted hyperboloid~$H_{(+)}$ for the positive mixtures, see fig.~1. Since these two surfaces~$H_{(\pm)}$ carry different topologies ($H_{(-)}$ is infinitely connected, the connected components of~$H_{(+)}$ are simply connected), a positive-mixture configuration can never be continously deformed into a negative mixture (and vice versa). This may be considered as the RST counterpart of the fermion-boson superselection rule of the conventional theory which is of purely phenomenological character.

The 2-particle configurations admit a concrete geometric realization if the underlying space-time manifold (as the base space of the applied fibre bundles) is parallelisable. As is well-known in differential geometry, any parallelisable manifold admits a {\it global} frame of tangent vectors and thus admits the introduction of a {\it flat} connection for its tangent bundle (though this does not imply in general the flatness of the tangent metric \cite{GrKlMay,Hi}). On the other hand, the RST 2-particle mixtures require the existence of a trivial principal Lorentz bundle over space-time, whose sections generate a flat Lorentz connection via the (pullback of the) Maurer-Cartan form over the Lorentz group $SO(1,3)$. The reduction of this 4-frame to a 2-frame then generates the bundle geometry of the 2-particle mixtures ({\bf Section~VI}).
\newpage
\section{Relativistic Schr\"odinger Theory}

In order to present the subsequent discussions in a sufficiently self-contained form, we briefly sketch the RST fundamentals with emphasis of those aspects which lead to the emergence of the matter dichotomy, the main subject of the present paper. According  to the RST claim of representing a general theory of matter, we evoke the mathematical framework of fibre bundles over the pseudo-Riemannian space-time as the base space and we identify the corresponding bundle connection with the physical gauge potentials according to the principle of minimal coupling.

Thus the coordinate-covariant derivative~$\bs{\nabla}$ refers to the Levi-Civita connection~~$\bs{\Gamma}$ of the pseudo-Riemannian fibre metric~$\bf{g}$ in the tangent bundle of space-time, the gauge potential~~$\bs{\A}$ acts as the bundle connection in the complex vector bundles over space-time (with typical fibre~$\CC^N$). The gauge-plus-coordinate covariant derivative of a bundle section is thus~$\bs{\D}=\bs{\nabla}+\bs{\A}$. The N-component wave function~$\Psi$ represents a section of the vector bundle and the intensity matrix~$\I$ (generalizing the concept of wave function) is a Hermitian section~$(\I=\bar{\I})$ of the product bundle whose typical fibre is the tensor product $\CC^N \otimes \overline{\CC^N}$ of the vector fibres~$\CC^N$. Further examples of operator-valued sections are the anti-Hermitian field strength~$\F_{\mu\nu}\,(=-\bar{\F}_{\mu\nu})$ as the curvature of the connection~$\A_{\mu}\,(=-\bar{\A}_{\mu})$ 
\begin{equation}
  \label{eq:Fcur}
  \F_{\mu\nu} = \nabla_\mu\A_\nu - \nabla_\nu\A_\mu + \left[\A_\mu,\A_\nu\right] \; .
\end{equation}
The Hamiltonian~$\H_{\mu}$ (being neither Hermitian nor anti-Hermitian) may be split up into its (anti-) Hermitian parts as follows
\begin{equation}
  \label{eq:Hsplit}
  \H_\mu = \hbar c \left( \K_\mu + i\L_\mu\right)\ .
\end{equation}
Here, both the kinetic field~$\K_\mu$ and the localization field~$\L_\mu$ are Hermitian  $(\K_\mu=\bar{\K}_\mu,\L_\mu=\bar{\L}_\mu)$ and build up the (anti-)Hermitian constituents of the Hamiltonian~$\H_\mu$.

\subsection{RST Dynamics}

The basic  dynamical equations of the theory may be considered here in a purely formal view as mathematically consistent links between those RST objects mentioned above. The first equation to be mentioned is the relativistic Schr\"odinger equation (RSE) for the wave function~$\Psi$ 
\begin{equation}
  \label{eq:RSE}
  i\hbar c \D_\mu\Psi = \H_\mu\Psi\ ,
\end{equation}
or its generalization, resp., the relativistic von Neumann equation (RNE) for the intensity matrix~$\I$:
  \begin{equation}
    \label{eq:calI}
    \D_{\mu} \I = \frac{i}{\hbar c} \left[\I\cdot\bar{\H}_\mu - \H_\mu\cdot\I \right]\ .
  \end{equation}
Contrary to Schr\"odinger's non-relativistic quantum mechanics, the relativistic Hamiltonian~$\H_\mu$ is here not a rigid object, to be fixed a priori, but is a dynamical variable obeying its own field equations. The first of these is the ``{\it integrability condition}''
\begin{equation}
  \label{eq:IntCond}
  \D_\mu\H_\nu-\D_\nu\H_\mu + \frac{i}{\hbar c}\left[\H_\mu,\H_\nu \right]=i\hbar c \F_{\mu\nu}
\end{equation}
and the second is the ``{\it conservation equation}''
\begin{equation}
  \label{eq:ConEq}
  \D^\mu\H_\mu- \frac{i}{\hbar c}\H^\mu \cdot \H_\mu = -i\hbar c\left(\frac{\M c}{\hbar} \right)^2
\end{equation}
with~$\M$ being the mass operator, i.e.\ for identical particles of mass M
\begin{equation}
  \label{eq:mass}
  \M \rightarrow M\cdot\bf{1} \; .
\end{equation}

\subsection{Conservation Laws}

The meaning of the integrability condition~\rf{eq:IntCond} is to ensure the existence of (local) solutions for the RST dynamics, but the conservation equation~\rf{eq:ConEq} implies the existence of certain conservation laws, such as the charge conservation for~$N$ particles~$(a=1\ldots N)$
\begin{equation}
  \label{eq:chcon}
  \nabla^\mu j_{a\mu} = 0\ ,
\end{equation}
or the energy-momentum law
\begin{equation}
  \label{eq:enlaw}
  \nabla^\mu T_{\mu\nu} = \f_\nu\ .
\end{equation}
Here the Lorentz force density~$\f_\nu$ is being composed of the field strength~$\F_{\mu\nu}=\{F_{a\mu\nu}\}$ and the currents~$\J_\mu=\{j_{a\mu}\}$ as usual
\begin{equation}
  \label{eq:ftrIf}
  \f_\nu = \tr\left(\I \cdot \ff_\nu \right)=\hbar c  F_{a \mu \nu} j^{a \mu}
\end{equation}
where the (Hermitian) force operator~$\ff_\nu$ is given by~\cite{MaRuSo} 
\begin{equation}
  \label{eq:fop}
  \ff_\nu = i\,\frac{\hbar}{2M}\left(\bar{\H}^\mu\cdot\F_{\mu\nu}+\F_{\mu\nu}\cdot\H^\mu
  \right)\ .
\end{equation}
Clearly if the energy-momentum content of the gauge field is included in the energy-momentum density~$T_{\mu\nu}$, one has the usual energy-momentum conservation
law~\cite{MaRuSo} for the total system
\begin{equation}
  \label{eq:EnCon}
  \nabla^\mu T_{\mu\nu} = 0\ .
\end{equation}

The last equation, closing the whole dynamical system, is taken as the (generally non-Abelian) Maxwell equation
\begin{equation}
  \label{eq:DFJ}
  \D^\mu \F_{\mu\nu} = 4\pi\alpha_* \J_\nu\ 
\end{equation}
\centerline{($\alpha_{\ast}=\frac{e^2}{\hbar c}$).}
Thus in the last step, it remains to specify the current operator~$\J_\mu$ in terms of the matter variable~$\I$ and then one has a closed dynamical system. The desired relationship between the intensity matrix~$\I$ and current~$\J_\mu$ is obtained by decomposing the current operator~$\J_\mu$ with respect to the generators~$\tau^a$ of the gauge group
\begin{equation}
  \label{eq:Jtau}
  \J_\mu = j_{a\mu}\tau^a
\end{equation}
and then defining the current components~$j_{a\mu}$ as
\begin{equation}
  \label{eq:jtrIv}
  j_{a\mu} = \tr\left(\I\cdot v_{a\mu} \right)
\end{equation}
where the velocity operators read in terms of the Hamiltonian~$\H_\mu$ for scalar particles
\begin{equation}
  \label{eq:vtauH}
  v_{a\mu}=\frac{i}{2Mc^2}\left(\tau_a\cdot\H_\mu+\bar{\H}_\mu\cdot\tau_a \right)
\end{equation}
(the velocity operator for a spinning particle coincides with the Dirac matrices:~$v_\mu\rightarrow\gamma_\mu$~\cite{MaSo99}). One can easily show that this arrangement automatically guarantees the validity of the charge conservation laws~\rf{eq:chcon} as a consequence of the RST dynamics with no need of additional requirements.

Such a pleasant effect occurs also for the energy-momentum laws~\rf{eq:enlaw} and~\rf{eq:EnCon}. Indeed it is possible to define the energy-momentum content of matter~$T_{\mu\nu}$ in such a way that the laws~\rf{eq:enlaw} and~\rf{eq:EnCon} are {\it automatically} obeyed again as an implication of the RST dynamics itself. More concretely, one defines the energy momentum density~$T_{\mu\nu}$ as
\begin{equation}
  \label{eq:TtrIT}
  T_{\mu\nu} = \tr \left(\I\cdot\T_{\mu\nu} \right)
\end{equation}
with the energy-momentum operator~$\T_{\mu\nu}$ for scalar particles being given through
\begin{equation}
  \label{eq:Tscp}
  \T_{\mu\nu}=\frac{1}{2Mc^2}\left[\bar{\H}_\mu\cdot\H_\nu+\bar{\H}_\nu\cdot\H_\mu
  -g_{\mu\nu}\left(\bar{\H}^\lambda \cdot \H_\lambda-\left(\M c^2 \right)^2 \right)\right]\ ,
\end{equation}
(see ref.~\cite{MaSo99} for the energy-momentum operator of Dirac particles). If the gravitational interactions are to be included, one adds the Einstein equations to the RST dynamics which however enforces the energy-momentum conservation law~\rf{eq:EnCon} as a consistency condition, in place of  the weaker source equation~\rf{eq:enlaw}. The stronger condition~\rf{eq:EnCon} is attained by adding the energy-momentum density of the gauge field to that of the matter field~\rf{eq:TtrIT} which then yields a closed system with the strong law~\rf{eq:EnCon} playing the part of a closedness condition~\cite{OcSo}.

\subsection{Amplitude Field}

The preceding examples of \emph{automatic} conservation laws have been presented in such a great detail because, for the present purpose of demonstrating the emergence of a RST matter dichotomy, it is important to have a further such automatic consequence of the RST dynamics: namely the emergence of a scalar amplitude field~$L(x)$ from the localization operator~$\L_{\mu}(x)$~\rf{eq:Hsplit}.

In order to see more clearly the reason why the general RST dynamics gives rise to such an amplitude scalar~$L(x)$, one first transcribes the integrability condition~\rf{eq:IntCond} to the kinetic and localization fields~$\K_\mu, \L_\mu$~\rf{eq:Hsplit} and thus finds for the Hermitian part~$\K_\mu$ of $\H_\mu$
\begin{equation}
  \label{eq:Kherm}
  \D_\mu\K_\nu-  \D_\nu\K_\mu + i\left[\K_\mu ,\K_\nu\right]-i \left[\L_\mu ,\L_\nu\right] =
  i\F_{\mu\nu} \; ,
\end{equation}
and similarly for the localization field~$\L_\mu$
\begin{equation}
  \label{eq:Lherm}
  \D_\mu\L_\nu-  \D_\nu\L_\mu + i \left[\L_\mu ,\K_\nu\right]+ i \left[\K_\mu ,\L_\nu\right] = 0 \; .
\end{equation}
By inspecting the trace of the localization field
\begin{equation}
  \label{eq:trL}
  L_\mu \doteqdot \tr\L_\mu \; ,
\end{equation}
one finds for the trace~$L_\mu$ of~$\L_\mu$ the following curl relation from (\ref{eq:Lherm}):
\begin{equation}
  \label{eq:curlL}
  \nabla_\mu L_\nu -   \nabla_\nu L_\mu = 0\ .
\end{equation}
Consequently, this vector field~$L_\mu$ must be a gradient field generated by some scalar ($L^N$, say)
\begin{equation}
  \label{eq:Lscal}
  L_\mu = \frac{\partial_\mu L^N}{L^N} = N\,\frac{\partial_\mu L}{L}\ .
\end{equation}
Here,~$N$ is chosen as the dimension of the typical vector fibre (i.e.~$N=2$ for two-particle systems, see below). The significance of this \emph{amplitude field}~$L(x)$ for the matter dichotomy will readily become evident.

\newpage
\section{Matter Dichotomy}

The very general features of RST mentioned up to now are already sufficient in order to conclude that the RST matter emerges in two different forms which are dynamically kept apart: the positive and negative mixtures, being separated by the pure states as a certain kind of borderline configuration. Indeed, as will become evident through the following arguments, we even need only the RNE~\rf{eq:calI} together with the integrability condition~\rf{eq:IntCond}in the form~\rf{eq:curlL}-\rf{eq:Lscal} in order to establish the desired matter dichotomy. This will be readily proven explicitly for the lowest-order cases~$N=2,3,4$ and from here it becomes evident by means of some inductive arguments that the dichotomy must exist for any fibre dimension~$N$ (which is the minimal number of fibre dimensions necessary for the description of $N$ scalar particles). 

\subsection{Deviators}

The point of departure is the recursive construction of certain polynomials~$\p{n}$ of degree~$(n+1)$ for the intensity matrix~$\I$. Introducing here the \emph{deviation densities}~$\Delta(n)$ as the trace of the {\it deviators} ~$\p{n}$, i.e.
\begin{equation}
  \label{eq:Deltan}
  \Delta(n)=\frac{1}{n}\tr\p{n}\ ,
\end{equation}
the deviators~$\p{n}$ are defined through
\begin{gather}
  \label{eq:Devi}
  \p{n} = \left(\Delta(n-1)\cdot{\bf 1}-\p{n-1} \right)\cdot\I \\*
  \big( n = 2,3,4,5,\ldots ). \notag
\end{gather}
The starting point of the recursion is the lowest-order deviator~$\p{1}$ which we put identical to the intensity matrix~$\I$ itself
\begin{subequations}
  \label{eq:D1}
\begin{align}
  \p{1} &\equiv \I \\
  \Delta(1) &\equiv \tr\I \doteqdot \rho \; .
\end{align}
\end{subequations}
Then comes the first (non-trivial) deviator~$(n=2)$
\begin{gather}
  \label{D2}
  \p{2} \doteqdot \DF = \rho\cdot\I - \I^2\\*
  \big(\DeF \doteqdot \Delta(2) = \frac{1}{2} \tr \DF \big)\ ;\notag
\end{gather}
correspondingly the second deviator reads~$(n=3)$
\begin{equation}
  \label{eq:D3}
  \begin{split}
    \p{3} \doteqdot \DS &= \left(\Delta(2)\cdot{\bf 1} - \p{2} \right)\cdot\I\\*
    &= \DeF\cdot\I - \DF\cdot\I = \I^3 - \rho\cdot\I^2 + \DeF\cdot\I\ \; .
  \end{split}
\end{equation}
And finally let us mention the third deviator~$\DT\,(\doteqdot\p{4})$
 \begin{gather}
  \label{eq:D4}
  \begin{split}
    \DT &= \left(\DeS\cdot{\bf 1} - \DS\right)\cdot\I\\*
    &= -\I^4+\rho\cdot\I^3-\DeF\cdot\I^2+\DeS\cdot\I
  \end{split}\\*
  \big( \DeS \doteqdot \frac{1}{3}\tr\DS\big)\notag\ .
 \end{gather}

It should be evident that this recursive construction of deviators becomes trivial when the deviation order~$n$ coincides with the fibre dimension~$N$, i.e.\ we have
\begin{equation}
  \label{eq:Dn0}
  \p{n} = 0\ ,\ n = N+1,N+2,N+3,\ldots
\end{equation}
Thus the last non-trivial deviator for fibre-dimension~$N$ is~$\p{N}$ and the next one vanishes, ~$\p{N+1}=0$, i.e.\ by means of equation~\rf{eq:Devi}
\begin{equation}
  \label{eq:DN0}
  \left(\Delta(N)\cdot{\bf 1} - \p{N}\right)\cdot\I = 0\ .
\end{equation}
However, for a regular intensity matrix~$(\det\I\ne 0)$ this implies for fibre dimension~$N$
\begin{equation}
  \label{eq:DN}
  \p{N}=\Delta(N)\cdot {\bf 1}\ ,
\end{equation}
which is nothing else than the well-known \emph{Hamilton-Cayley identity} of matrix calculus~\cite{La}. Thus for fibre dimension~$N=2$ we conclude for the~$(2\times2)$-matrix~$\I$
\begin{equation}
  \label{eq:DF}
  \DF\equiv \rho\I-\I^2 = \DeF\cdot{\bf 1}\ ,
\end{equation}
with the deviation density~$\DeF$ being identical to the determinant of the intensity matrix, i.e.
\begin{equation}
  \label{eq:detI2}
  \DeF = \det\I\ ,\ (N=2)\ .
\end{equation}
Similarly for the fibre dimension~$N=3$ one deduces from equation~\rf{eq:D3} for the~$3\times 3$-matrix~$\I$
\begin{equation}
  \label{eq:DS3}
  \DS\equiv\DeF\cdot\I-\rho\I^2+\I^3 = \DeS\cdot{\bf 1}\ ,
\end{equation}
where the second deviation density~$\DeS$ is now the determinant of the intensity matrix~$\I$
\begin{equation}
  \label{eq:detI3}
  \DeS = \det\I\ ,\ (N=3)\ .
\end{equation}
Or finally, for fibre dimension~$N=4$, the closing condition~\rf{eq:DN} yields on account of equation~\rf{eq:D4} the corresponding Hamilton-Cayley identity
\begin{equation}
  \label{eq:DT}
  \DT=\DeS\cdot\I-\DeF\I^2+\rho\I^3-\I^4 \equiv \DeT\cdot{\bf 1}
\end{equation}
with
\begin{equation}
  \label{eq:DeT}
  \DeT = \det\I\  ,\ (N=4)\ .
\end{equation}

These few examples (in place of a rigorous induction scheme) may be sufficient in order to see the general formalism working well for any fibre dimension~$N$. It will be used now in order to establish the desired dichotomy effect.

\subsection{Matter Dichotomy}

The interesting point with the RST dynamics is now that it subdivides the configuration space of matter (i.e.\ the space of regular~$(N\times N)$-matrices) into two dynamically separated subsets: the positive and negative mixtures, with the pure states appearing as an asymptotic limit of both kinds of mixtures. Here the criterium of membership refers to the determinant~$\det\I=\Delta(N)$ of the intensity matrix~$\I$ such that for~$\Delta(N)>0\ (\Delta(N)<0)$ one has a ``positive'' (``negative'') mixture. On the other hand the pure states are uniquely characterized by the well-known Fierz identity~\cite{MaSo99} for the intensity matrix, i.e.
\begin{equation}
  \label{eq:DF0}
  \DF\equiv 0 \; ,
\end{equation}
which restricts~$\I$ to the tensor product of some wave function~$\Psi$:
\begin{equation}
  \label{eq:IPsi}
  \I\rightarrow\Psi\otimes\bar{\Psi}\ .
\end{equation}
Indeed, this is easily verified by simply substituting the latter form of~$\I$ into the definition of the first deviator~$\DF$~\rf{D2}. For the special case of two fibre dimensions $N=2$  the Hamilton-Cayley identity~\rf{eq:DF} says that the Fierz identity upon~$\I$~\rf{eq:DF0} is even equivalent to the one real condition of vanishing Fierz deviation~$\DeF=0$ (see below for a detailed treatment of the two-particle systems).

However, by what circumstance does this dichotomic mixture effect come about? The answer comes from the combination of the RNE~\rf{eq:calI} with the existence of the amplitude field~$L$~\rf{eq:Lscal}. Here, the first step consists in splitting up the $N$-order deviator~$\p{N}$ into the highest-power term of~$\I$ and the remainder~$\Y(N)$:
\begin{equation}
  \label{eq:DNY}
  \p{N} = - (-1)^N\cdot\I^N + \Y(N)\ .
\end{equation}
For instance from the definition of the first deviator~$\DF$~\rf{D2} we obtain
\begin{equation}
  \label{eq:Y2}
  \Y(2) \doteqdot \Y_F = \rho\I \; ,
\end{equation}
or similarly for the second deviator~$\DS$~\rf{eq:D3}
\begin{equation}
  \label{eq:Y3}
  \Y(3) \doteqdot \Y_S = \rho\DF + \left(\DeF-\rho^2 \right)\I
\end{equation}
and finally for the third deviator~$\DT$~\rf{eq:D4}
\begin{equation}
  \label{eq:Y4}
  \Y(4) \doteqdot \Y_T = \rho\cdot\DS + \left(\DeF-\rho^2\right)\cdot\DF + \left(\rho^3-2\rho\DeF +
    \DeS\right)\cdot\I\ .
\end{equation}
The general procedure should be self-suggesting  from these few examples.

In the next step, one considers the derivative of the determinant~$\Delta(N)$~\rf{eq:Deltan} which is found to consist of two terms according to the splitting~\rf{eq:DNY}
\begin{equation}
  \label{eq:split}
\begin{split}
  \partial_\mu\Delta(N) &= \frac{1}{N}\tr\left(\D_\mu \p{N}\right)\\*
  &= 2\tr\left(\p{N}\cdot\L_\mu\right)+\frac{1}{N}\tr\left(\D_\mu \Y(N) - 2N
    \Y(N)\cdot\L_\mu\right)\ .
\end{split}  
\end{equation}
The first term on the right is due to the derivative of the first term~$(\sim \I^N)$ of the splitting~\rf{eq:DNY} when the RNE~\rf{eq:calI} has been used and after this that splitting~\rf{eq:DNY} is applied again in the opposite direction leading back to~$\p{N}$. However since~$\p{N}$ is of that simple form~\rf{eq:DN}, namely proportional to the identity~${\bf{1}}(N)$, the derivative of the determinant~$\Delta(N)$~\rf{eq:split} becomes
\begin{equation}
  \label{eq:dDeltaN}
  \partial_\mu\Delta(N) = 2\Delta(N)\cdot\tr\L_\mu
\end{equation}
provided the second term on the right of~\rf{eq:split} vanishes, i.e. concretely if we have the trace identity
\begin{equation}
  \label{eq:trid}
  \tr\left(\D_\mu \Y(N) - 2N\Y(N)\cdot\L_\mu\right) \equiv 0 \ .
\end{equation}

But this identity actually does hold, from which one is easily convinced by the above mentioned examples~$(N=2,3,4)$ and the corresponding induction arguments. Indeed, one merely observes for the first term on the left of~\rf{eq:trid}
\begin{equation}
  \label{eq:trDY}
  \tr\left(\D_\mu \Y(N)\right) = \partial_\mu\left(\tr \Y(N)\right)
\end{equation}
with the trace of~$\Y(N)$ being easily computed by means of the definition of the deviation densities~$\Delta(N)$~\rf{eq:Deltan} for~$n \le N$. Thus for~$N=4$, e.g., one immediately finds from equation~\rf{eq:Y4}
\begin{equation}
  \label{eq:trDY2}
  \tr\left(\D_\mu\Y_T\right)=\partial_\mu\left(4\rho\DeS-4\rho^2\DeF+2\DeF^2+\rho^4\right)\
  .
\end{equation}
On the other hand, the computation of the second term on the left of~\rf{eq:trid} just meets with this result because for the lower-order deviation densities~$\Delta(n)\ (n\le N)$ one also has
\begin{equation}
  \label{eq:trDnL}
  \tr\left(\p{n}\cdot\L_\mu\right) = \frac{1}{2} \partial_\mu\Delta(n)\ ,
\end{equation}
cf.~\rf{eq:split}, and this then validates definitively the trace identity~\rf{eq:trid}. Observe here that the latter relation~\rf{eq:trDnL}, though being seen now to hold for all deviation orders~$n$ up to the maximal value~$n=N$, had to be exploited only for the maximal order~$N$ (in order to yield the desired result~\rf{eq:dDeltaN} because the closing relation~\rf{eq:DN} does apply exclusively to this maximal order~$N$!

Once the decisive derivative relation~\rf{eq:dDeltaN} is now firmly established, the matter dichotomy emerges by simply observing the former gradient property for the location field~$\L_\mu$~\rf{eq:trL}-\rf{eq:Lscal}. Indeed the existence of the amplitude field~$L(x)$~\rf{eq:Lscal} admits to formally integrate the differential equation~\rf{eq:dDeltaN} for~$\Delta(N)$ to yield
\begin{equation}
  \label{eq:DNdetIN}
  \Delta(N)\equiv\det\I(N)= \Din\left(\frac{L(x)}{\Lin} \right)^{2N}\ .
\end{equation}
Thus, at any event~$x$ of space-time the determinant of the intensity matrix~$\I$ has the same sign as at the initial event~$\xin$ provided both events~$x$ and~$\xin$ can be connected by some path along which the amplitude field~$L(x)$ is a regular scalar. The conclusion is that whenever the RST dynamics admits a regular solution over some region of space-time, then this field configuration must be either a positive mixture~$(\Din>0)$ or a negative mixture~$(\Din<0)$ or a "quasi-pure state"~$(\Din=0)$. The latter kind of mixtures are not ordinary pure states (which have~$\DF=0$, see~\rf{eq:DF0}) but they are not forbidden by the RST dynamics to decay to such an ordinary pure state (with vanishing deviator~$\DF$); however the positive and negative mixtures must strictly preserve their mixture character over the whole domain of regularity in space-time (\emph{"matter dichotomy"}).

\newpage
\section{Two-Particle Systems}

When matter appears in two different forms, one clearly wants to see the common features and also the differences. For a study of such questions in the subsequent sections, we restrict ourselves to the two-particle systems ($N=2$) and we specialize first the general dynamical equations for this situation. Here it is important to parametrize the RST field system in such a way that a clear separation between single-particle variables and exchange fields comes about. Such a conceptual separation of the two field subsystems facilitates afterwards the discussion of the exchange effects occuring between the single-particle subsystems. 
\subsection{Operator Kinematics}
As attractive as the covariant equations may look like in their abstract form, for concrete computations one mostly resorts to a special reference system and decomposes the covariant objects with respect to such a reference frame. The corresponding components of the (gauge and/or coordinate) covariant objects are then treated as the proper dynamical objects entering the component form of the abstract dynamical equations. For the present RST operator dynamics, one thus chooses first some orthonormal basis for the operators acting over the typical  vector fibre ${\mathbb{C}}^2$ of the two-particle systems. This may be done by choosing two {\it single-particle projectors} $\P_a$ ($a,b|=1,2$)   
\begin{subequations}
\label{e4:1} 
\begin{align}
\P_a \cdot \P_b &= \delta_{ab} \cdot \P_a \\
\P_a &=\bar{\P}_a\\
\P_1 + \P_2 &= {\bf 1}\\
tr \, \P_a &=1 \; , 
\end{align}
\end{subequations}
and by complementing these by two {\it permutators} $\hat{\Pi}$, $\tilde{\Pi}$ which obey the following algebraic relations:
\begin{subequations}
\label{e4:2} 
\begin{align}
\hat{\Pi} \cdot \hat{\Pi} &= \tilde{\Pi} \cdot \tilde{\Pi}={\bf 1} \\
\big\{\hat{\Pi},\tilde{\Pi} \big\} &=0 \\
\big[\hat{\Pi},\tilde{\Pi} \big] &=2i(\P_1-\P_2) \\
\big\{\hat{\Pi},\P_a \big\} & = -i\big[\tilde{\Pi},\P_1\big]=i\big[\tilde{\Pi},\P_2\big]=\hat{\Pi} \\
\big\{\tilde{\Pi},\P_a \big\} & =- i\big[\hat{\Pi},\P_1\big]=i\big[\hat{\Pi},\P_2\big]=\tilde{\Pi} \; .  
\end{align}
\end{subequations}
The four (Hermitian) operators $\{\P_a$, $\hat{\Pi}$, $\tilde{\Pi}\}$ form a complete basis over the 2-particle fibre ${\mathbb{C}}^2$ and consequently the generators $\tau_a$ (\ref{eq:Jtau}) of the gauge group must be certain linear combinations in this basis set; for the (abelian) electromagnetic interactions one has \cite{MaRuSo}
\begin{equation}
\label{e4:3}
\tau_a=-i\P_a
\end{equation}
and thus the (anti-Hermitian) bundle curvature $\F_{\mu \nu}(=-\bar{\F}_{\mu \nu})$ and its connection $\A_{\mu}(=-\bar{\A}_{\mu})$ are decomposed as follows
\begin{subequations}
\label{e4:4} 
\begin{align}
\F_{\mu \nu} &= F_{a \mu \nu}\, \tau^a \\
\A_{\mu} &= A_{a \mu}\, \tau^a \; .
\end{align}
\end{subequations}

Once the reference section for the operators has been fixed, any operator-valued section can be decomposed with respect to this {\it rotating basis} $\{\P_a$, $\hat{\Pi}$, $\tilde{\Pi}\}$, e.g. for the intensity matrix $\I$
\begin{equation}
\label{e4:5}
\I=\rho_a \P^a +\frac{1}{2}s \, \hat{\Pi} \; .
\end{equation}
Properly speaking, any Hermitian ($2\times 2$)-matrix (such as $\I$) in general has $2 \times 2=4$ components but the rotating basis is co-moving with the intensity matrix in such a way that one of the four components of $\I$ is kept zero permanently. Due to the orthonormality of the rotating operator basis the components $\rho_a$, $s$ of $\I$ can be extracted from $\I$ in the usual way applicable to any Hilbert-space structure, namely the {\it single-particle densities} $\rho_a$ as 
\begin{equation}
\label{e4:6}
\rho_a=tr (\P_a \cdot \I)
\end{equation}
and the {\it overlap density} $s$ as
\begin{equation}
\label{e4:7}
s=tr (\hat{\Pi} \cdot \I) \; .
\end{equation}
Similarly, the decomposition of the kinetic field $\K_{\mu}$ (\ref{eq:Hsplit}) reads
\begin{equation}
\label{e4:8}
\K_{\mu}=K_{a \mu}\cdot \P^a + \paraQ_{\mu} \cdot \hat{\Pi} + \perpQ_{\mu} \cdot \tilde{\Pi} \; ,
\end{equation}
or for the localization field $\L_{\mu}$
\begin{equation}
\label{e4:9}
\L_{\mu}=L_{a \mu}\cdot \P^a + \paraN_{\mu} \cdot \hat{\Pi} + \perpN_{\mu} \cdot \tilde{\Pi} \; .
\end{equation}
The coefficients $K_{a \mu}$, $L_{a \mu}$ in front of the projectors $\P_a$ are the {\it single-particle} components and correspondingly the coefficient fields in front of the permutators $\hat{\Pi}$, $\tilde{\Pi}$ are the {\it exchange fields}. From (\ref{e4:9}) one immediately recognizes the former amplitude vector $L_{\mu}$ (\ref{eq:trL}) as the sum of the corresponding single-particle vectors
\begin{equation}
\label{e4:10}
L_{\mu}=tr \, \L_{\mu} = L_{1 \mu} +L_{2 \mu} \; . 
\end{equation}

One could now write down the whole RST dynamics (\ref{eq:RSE})-(\ref{eq:ConEq}) in component form referring to those component fields just introduced through (\ref{e4:5})-(\ref{e4:9}) but for the interpretation of the physical effects it is very favorable to pass over to a new combination of these component fields. 

\subsection{Reparametrization}
The first hint upon the advantage of dealing with a new combination of the exchange fields $N_{\mu}, \; Q_{\mu}$ (\ref{e4:8})-(\ref{e4:9}) comes from the density dynamics, i.e. the component form of the RNE (\ref{eq:calI})
\begin{subequations}
\label{e4:11} 
\begin{align}
\partial_{\mu} \rho &=\rho \cdot L_{\mu} + q \cdot l_{\mu} + 2s \cdot \paraN_{\mu} \\
\partial_{\mu} q &=\rho \cdot l_{\mu} +q \cdot L_{\mu} - 2s \cdot \perpQ_{\mu}\\
\partial_{\mu} s &= s\cdot L_{\mu} +2\rho\cdot \paraN_{\mu} +2q \cdot \perpQ_{\mu} \; . 
\end{align}
\end{subequations}
Here we have made use of the total density $\rho$ (\ref{eq:D1}b) as the trace of the intensity matrix $\I$ (\ref{e4:5})
\begin{equation}
\label{e4:12}
\rho=tr \, \I=\rho_1+\rho_2
\end{equation}
and also of the relative density
\begin{equation}
\label{e4:13}
q=\rho_1-\rho_2 \; ,
\end{equation}
as well as the difference ($l_{\mu}$) of localization fields $L_{a \mu}$
\begin{equation}
\label{e4:14}
l_{\mu}=L_{1 \mu}-L_{2 \mu} \; .
\end{equation}

The interesting point with the density system (\ref{e4:11}) is now that it admits a conserved quantity ($\sigma_{\ast}$, say) emerging from the following relation which is a consequence of that system: 
\begin{equation}
\label{e4:15}
\partial_{\mu}\,(\rho^2-q^2-s^2)=2L_{\mu}\,(\rho^2-q^2-s^2) \; .
\end{equation}
This may be formally integrated to yield 
\begin{equation}
\label{e4:16}
(\rho^2-q^2-s^2){(x)}=(\rho^2-q^2-s^2)_{in}\left(\frac{L(x)}{L_{in}}\right)^4 \; .
\end{equation}
Clearly this result is nothing else than the 2-particle specialization ($N=2$) of the general result (\ref{eq:DNdetIN}), since for two fibre dimensions the first deviation density $\Delta_F$ (\ref{D2}) is found as
\begin{equation}
\label{e4:17}
\Delta_F=det\,\I=\frac{1}{2} tr\, \D_F=\frac{1}{4}(\rho^2-q^2-s^2) \; .
\end{equation}
In this way it is once more seen very clearly that the emergence of the RST matter dichotomy is an immediate consequence of the dynamical equations, not a kinematical postulate as in the conventional quantum theory.

Now the latter result (\ref{e4:16}) suggests to reparametrize the densities $\rho$, $q$, $s$ by means of the amplitude field $L(x)$ and certain renormalization factors $Z_T$, $Z_R$, $Z_O$ as follows
\begin{subequations}
\label{e4:18} 
\begin{align}
\rho &=Z_T \cdot L^2 \\
q &=Z_R \cdot L^2 \\
s &=Z_O \cdot L^2 \; .
\end{align}
\end{subequations}
As a consequence of this arrangement the following constraint upon the renormalization factors is obtained
\begin{equation}
\label{e4:19}
Z_T^2-Z_R^2-Z_O^2=\sigma_{\ast} 
\end{equation}
with the {\it mixing index} $\sigma_{\ast}$ taking the values: $\sigma_{\ast}=+1$ (positive mixtures), $\sigma_{\ast}=0$ (pure states), and $\sigma_{\ast}=-1$ (negative mixtures) (fig.1). Thus, for such a small fibre dimension ($N=2$) the quasi-pure states coincide with the ordinary pure states, see the remarks below equation (\ref{eq:DNdetIN}). Clearly with that arrangement (\ref{e4:18}) for the renormalization factors, their dynamical equations are deduced from the original density dynamics (\ref{e4:11}) and are thus found to be of the following form:
\begin{subequations}
\label{e4:20} 
\begin{align}
\partial_{\mu} Z_T &= Z_R \cdot l_{\mu} + 2\, Z_O \cdot \paraN_{\mu} \\
\partial_{\mu} Z_R &= Z_T \cdot l_{\mu} - 2\, Z_O \cdot \perpQ_{\mu}\\
\partial_{\mu} Z_O &= 2 (Z_T \cdot \paraN_{\mu} + Z_R \cdot \perpQ_{\mu}) \; . 
\end{align}
\end{subequations} 

The second step of reparametrization refers now to the exchange fields $Q_{\mu}$, $N_{\mu}$. From the third equation (\ref{e4:20}c) of the renormalization dynamics one concludes that the special combination of exchange fields $\paraN_{\mu}$, $\perpQ_{\mu}$ emerging there on the right-hand side must be a gradient field ($S_{\mu}$, say):
\begin{equation}
\label{e4:21}
S_{\mu} \doteqdot \frac{1}{2} \frac{\partial_{\mu}Z_O}{Z_O} \; .
\end{equation}
This suggests to introduce a new combination ($S_{\mu}$, $T_{\mu}$) in place of the original ($N_{\mu}$, $Q_{\mu}$) in the following way:
\begin{subequations}
\label{e4:22} 
\begin{align}
S_{\mu} &= \frac{Z_T}{Z_O} \cdot \paraN_{\mu} + \frac{Z_R}{Z_O} \cdot \perpQ_{\mu} \\
T_{\mu} &= \frac{Z_R}{Z_O} \cdot \paraN_{\mu} + \frac{Z_T}{Z_O} \cdot \perpQ_{\mu}\; .
\end{align}
\end{subequations} 
With respect to these new exchange fields, the renormalization dynamics (\ref{e4:20}) adopts now the corresponding new form
\begin{subequations}
\label{e4:23} 
\begin{align}
\partial_{\mu} Z_T &= Z_R \cdot l_{\mu} + 2\lambda\,(Z_T\cdot S_{\mu} -Z_R\cdot T_{\mu}) \\
\partial_{\mu} Z_R &= Z_T \cdot l_{\mu} - 2\lambda\,(Z_T\cdot T_{\mu} -Z_R\cdot S_{\mu})\\
\partial_{\mu} Z_O &= 2\,Z_O \cdot S_{\mu} \; . 
\end{align}
\end{subequations}

This reparametrization embraces different kinds of mixtures via the new scalar field $\lambda(x)$ which is defined through
\begin{equation}
\label{e4:24}
\lambda \doteqdot \frac{1}{\displaystyle 1+\frac{\sigma_{\ast}}{Z_O{}^2}}
\end{equation}
and which obeys the field equation
\begin{equation}
\label{e4:25}
\partial_{\mu}\lambda =4\lambda(1-\lambda)\cdot S_{\mu} \; .
\end{equation}
From its definition (\ref{e4:24}) one immediately concludes that $\lambda < 1$ for the positive mixtures ($\sigma_{\ast}=+1$) and that $\lambda > 1$ for the negative mixtures ($\sigma_{\ast}=-1$) while $\lambda = 1$ for the pure states ($\sigma_{\ast}=0$). Since the different kinds of mixtures are kept separated for dynamical reasons (matter dichotomy), the field equation for the {\it mixture variable} $\lambda$ must have solutions which strictly respect this dynamical separation. However this requirement is easily seen to be satisfied because the new exchange field $S_{\mu}$ (\ref{e4:22}a) is found to be a gradient field, cf. (\ref{e4:21})
\begin{equation}
\label{e4:26}
S_{\mu}\equiv \frac{\partial_{\mu}S}{S} \; ,
\end{equation}
and thus the field equation for $\lambda$ (\ref{e4:25}) is easily integrated to yield
\begin{equation}
\label{e4:26_2}
\lambda=\quad\left\{\quad \begin{array}{c}\displaystyle \frac{S^4}{S^4+S_c{}^4} \quad\mbox {{\rm , $\sigma_{\ast}=+1$}} \\[3 mm] \displaystyle \frac{S^4}{S^4-S_c{}^4} \quad\mbox {{\rm , $\sigma_{\ast}=-1$}} \end{array} \right.  
\end{equation}
with $S_c$ being an arbitrary integration constant. Though this result merely reproduces the original definition of $\lambda$ (\ref{e4:24}) through the identification
\begin{equation}
\label{e4:27}
Z_O{}^2 =\left(\frac{S}{S_{c}}\right)^4\; ,
\end{equation}
the differential version of which is just the former equation (\ref{e4:21}), we thus can be sure that the matter dichotomy is correctly incorporated also into the dynamics of the new set of variables: the field equation for $\lambda$ (\ref{e4:25}) admits either solutions (\ref{e4:26_2}) with $\lambda<1$ ($\leadsto$ positive mixtures: $\sigma_{\ast}=+1$, $0<S^2<\infty$) or solutions with $\lambda>1$ ($\leadsto$ negative mixtures: $\sigma_{\ast}=-1$, $S_c{}^2<S^2<\infty$). This is the two-particle realization of the general result explained at the end of Section III, below equation (\ref{eq:DNdetIN}).  

A special situation is encountered for vanishing exchange fields, especially $S_{\mu} \equiv 0$. For this case, the new scalar field $S(x)$ (\ref{e4:26}) must be a constant over space-time ($S_{\ast}$, say) which then implies the constancy of the new variable $\lambda$ ($\lambda \Rightarrow \lambda_{\ast}={\rm const}$) through the present result (\ref{e4:26_2}) 
\begin{equation}
\label{e4:28}
\lambda_{\ast}=\frac{S_{\ast}{}^4}{S_{\ast}{}^4+\sigma_{\ast}\cdot S_c{}^4} \; .
\end{equation}
Thus, the vanishing of the exchange fields leaves us alone with the mixture configurations which then are found to be degenerate in the sense that they generate the same physics as the pure states ({\it mixture degeneracy} \cite{RuSo01}). In this case there exists a group of mixing transformations ({\it mixing group}) which admits to construct the mixture configurations from the pure states and the group parameter is just the present constant $\lambda_{\ast}$ (\ref{e4:28}). Thus we clearly recognize the effect of switching-on the exchange fields, namely to make the mixing group parameter space-time dependent: $\lambda_{\ast} \Rightarrow \lambda(x)$. The resulting cooperation of mixture and exchange effects is just the essence of the RST quantum effects (see a separate paper).

\subsection{Physical Densities}
The reparametrization of exchange fields (\ref{e4:22}) is not complete because it refers to only two of the four exchange fields. However a corresponding transformation for the remaining exchange fields $\paraQ_{\mu}$ and $\perpN_{\mu}$ suggests itself when we consider now the physical densities. 

Turning first to the charge densities $j_{a \mu}$ (\ref{eq:jtrIv}), one substitutes the decompositions of the intensity matrix $\I$ (\ref{e4:5}) and of the Hamiltonian $\H_{\mu}$ (\ref{e4:8})-(\ref{e4:9}) into that definition (\ref{eq:jtrIv}) for the currents $j_{a \mu}$ and then finds the following result \cite{MaRuSo}
\begin{subequations}
\label{e4:29} 
\begin{align}
j_{1 \mu} &=\frac{\hbar}{Mc} \Big\{ \rho_1 K_{1 \mu}+\frac{1}{2}s\,\Big(\paraQ_{\mu}+\perpN_{\mu}\Big) \Big\} \\
j_{2 \mu} &=\frac{\hbar}{Mc} \Big\{ \rho_2 K_{2 \mu}+\frac{1}{2}s\,\Big(\paraQ_{\mu}-\perpN_{\mu}\Big) \Big\} \; .
\end{align}
\end{subequations}
The specific shape of this result suggests now to introduce two new fields \{$F_{\mu}$,$G_{\mu}$\} in place of \{$\paraQ_{\mu}$, $\perpN_{\mu}$\}, namely in the following way
\begin{subequations}
\label{e4:30} 
\begin{align}
F_{\mu} &= \frac{Z_T}{Z_O}\cdot \paraQ_{\mu} - \frac{Z_R}{Z_O}\perpN_{\mu} \\
G_{\mu} &= \frac{Z_R}{Z_O}\cdot \paraQ_{\mu} - \frac{Z_T}{Z_O}\perpN_{\mu} \; .
\end{align}
\end{subequations}
The reason for this arrangement is because both currents $j_{a \mu}$ (\ref{e4:29}) adopt a very simple form, namely
\begin{subequations}
\label{e4:31} 
\begin{align}
j_{1 \mu} &=\frac{\hbar}{Mc}\rho_1 \Big\{ K_{1 \mu}+\lambda(F_{\mu}-G_{\mu}) \Big\} \\
j_{2 \mu} &=\frac{\hbar}{Mc}\rho_2 \Big\{ K_{2 \mu}+\lambda(F_{\mu}+G_{\mu}) \Big\} 
\end{align}
\end{subequations}
and this form even simplifies further to 
\begin{subequations}
\label{e4:32} 
\begin{align}
j_{1 \mu} &=\frac{\hbar}{Mc}\rho_1\, {\mathbb{K}}_{1 \mu} \\
j_{2 \mu} &=\frac{\hbar}{Mc}\rho_2\, {\mathbb{K}}_{2 \mu}
\end{align}
\end{subequations}
when one simultaneously introduces the {\it effective kinetic fields} ${\mathbb{K}}_{a \mu}$ ($a=1,2$) through
\begin{subequations}
\label{e4:33} 
\begin{align}
{\mathbb{K}}_{1 \mu} &=K_{1 \mu}+\lambda(F_{\mu}-G_{\mu})\\
{\mathbb{K}}_{2 \mu} &=K_{2 \mu}+\lambda(F_{\mu}+G_{\mu})\; .
\end{align}
\end{subequations}
Thus the final form of the currents $j_{a \mu}$ (\ref{e4:32}) is the same which is obtained also by putting all the exchange fields to zero in the original version (\ref{e4:29}) and replacing the original kinetic fields $K_{a \mu}$ by their effective counterparts ${\mathbb{K}}_{a \mu}$ (\ref{e4:33}). Such an absorption of the exchange fields into the single-particle fields in order to build up the effective fields is not only a purely formal advantage but yields also a deeper insight into the physical essence of the theory. For the special case of pure states ($\lambda = 1$), the present reparametrization of the exchange fields in terms of $S_{\mu}$, $T_{\mu}$, $F_{\mu}$, $G_{\mu}$ has been described in a preceding paper \cite{RuHuSo} where the new parametrization was helpful for a discussion of the exchange degeneracy.

Once the kinetic fields have been transcribed to a new form, it is strongly suggestive to try a similar transcription for the localization fields $L_{a \mu}$ (\ref{e4:9}). The precise form of the desired transformation to the  {\it effective localization fields} ${\mathbb{L}}_{a \mu}$ ($a=1,2$) comes from a closer inspection of the energy-momentum density $T_{\mu \nu}$ (\ref{eq:TtrIT}) which however proceeds along the same line of arguments as applied also to the currents $j_{a \mu}$. Thus one is led to introduce the effective localization fields ${\mathbb{L}}_{a \mu}$ through
\begin{subequations}
\label{e4:34} 
\begin{align}
{\mathbb{L}}_{1 \mu} &=L_{1 \mu}+\lambda(S_{\mu}-T_{\mu})\\
{\mathbb{L}}_{2 \mu} &=L_{2 \mu}+\lambda(S_{\mu}+T_{\mu})
\end{align}
\end{subequations}
and these new fields build then up the energy-momentum density $T_{\mu \nu}$, albeit in a somewhat more complicated way as was the case with the kinetic fields in connection with the currents $j_{a \mu}$ (\ref{e4:32}). In order to display the general structure of $T_{\mu \nu}$, it is convenient to split it up into three parts:
\begin{equation}
\label{e4:35}
T_{\mu \nu}={}^{(1)}T_{\mu \nu}+{}^{(2)}T_{\mu \nu}+{}^{(x)}T_{\mu \nu}
\end{equation}
where the first two terms ${}^{(a)}T_{\mu \nu}$ ($a=1,2$) are the single-particle contributions and the third term ${}^{(x)}T_{\mu \nu}$ contains the exchange and mixture effect. Similarly as for the currents $j_{a \mu}$, the single-particle contributions ${}^{(a)}T_{\mu \nu}$ do no longer contain here explicitly the exchange fields since these have been absorbed into the effective single-particle fields ${\mathbb{K}}_{a \mu}$, ${\mathbb{L}}_{a \mu}$:

\begin{subequations}
\label{e4:36} 
\begin{align}
{}^{(1)}T_{\mu \nu} =& \frac{\hbar^2}{M}\, \rho_1 \bigg\{ {\mathbb{K}}_{1 \mu}{\mathbb{K}}_{1 \nu}+{\mathbb{L}}_{1 \mu}{\mathbb{L}}_{1 \nu} \nonumber \\ 
& -\frac{1}{2}g_{\mu \nu} \bigg[{\mathbb{K}}_{1 \lambda}{\mathbb{K}}_1{}^{\lambda} +{\mathbb{L}}_{1 \lambda}{\mathbb{L}}_1{}^{\lambda} - \left(\frac{Mc}{\hbar} \right)^2 \bigg] \bigg\} \\[3 mm]
{}^{(2)}T_{\mu \nu} =& \frac{\hbar^2}{M}\, \rho_2 \bigg\{ {\mathbb{K}}_{2 \mu}{\mathbb{K}}_{2 \nu}+{\mathbb{L}}_{2 \mu}{\mathbb{L}}_{2 \nu} \nonumber \\
& -\frac{1}{2}g_{\mu \nu} \bigg[{\mathbb{K}}_{2 \lambda}{\mathbb{K}}_2{}^{\lambda} +{\mathbb{L}}_{2 \lambda}{\mathbb{L}}_2{}^{\lambda} - \left(\frac{Mc}{\hbar} \right)^2 \bigg] \bigg\} \; .
\end{align}
\end{subequations}
On the other hand, the exchange term ${}^{(x)}T_{\mu \nu}$ is built up itself from two contributions 
\begin{equation}
\label{e4:37}
{}^{(x)}T_{\mu \nu}={}^{(1)}t_{\mu \nu}+{}^{(2)}t_{\mu \nu}
\end{equation}
which do not contain the single-particle fields and thus look as follows
\begin{subequations}
\label{e4:38} 
\begin{align}
{}^{(1)}t_{\mu \nu} =& \frac{\hbar^2}{M}\, \lambda(1-\lambda)\rho_1 \Big\{ (F_{\mu}-G_{\mu})(F_{\nu}-G_{\nu}) +(S_{\mu}-T_{\mu})(S_{\nu}-T_{\nu}) \nonumber \\ 
& -\frac{1}{2}g_{\mu \nu} \Big[(F_{\lambda}-G_{\lambda})(F^{\lambda}-G^{\lambda}) +(S_{\lambda}-T_{\lambda})(S^{\lambda}-T^{\lambda}) \Big] \Big\} \\
{}^{(2)}t_{\mu \nu} =& \frac{\hbar^2}{M}\, \lambda(1-\lambda)\rho_2 \Big\{ (F_{\mu}+G_{\mu})(F_{\nu}+G_{\nu}) +(S_{\mu}+T_{\mu})(S_{\nu}+T_{\nu}) \nonumber \\[3 mm] 
& -\frac{1}{2}g_{\mu \nu} \Big[(F_{\lambda}+G_{\lambda})(F^{\lambda}+G^{\lambda}) +(S_{\lambda}+T_{\lambda})(S^{\lambda}+T^{\lambda}) \Big] \Big\} \; .
\end{align}
\end{subequations}

Observe here that, in contrast to the currents $j_{a \mu}$ (\ref{e4:32}), the energy-momentum density $T_{\mu \nu}$ (\ref{e4:35}) contains an extra term (namely ${}^{(x)}T_{\mu \nu}$) which even after the absorption of the exchange fields into the new single-particle fields ${\mathbb{K}}_{a \mu}$, ${\mathbb{L}}_{a \mu}$ is built up explicitly by all the exchange fields ($S_{\mu}$, $T_{\mu}$, $F_{\mu}$, $G_{\mu}$) together with the mixture variable $\lambda(x)$. Such a term in $T_{\mu \nu}$ is necessary in order to validate the energy-momentum conservation law (\ref{eq:EnCon}) (or (\ref{eq:enlaw}), resp.) whereas the charge conservation law (\ref{eq:chcon}) works without such an {\it explicit} exchange term. Observe also that the latter term ${}^{(x)}T_{\mu \nu}$ is built up by {\it both} the mixture variable $\lambda(x)$ {\it and} the exchange fields and thus represents the cooperation of both kinds of field degrees of freedom. 

\subsection{Hamiltonian Dynamics}
If one wishes to convince oneself of the consistency of those conservation laws for the physical densities, one has to use the source and curl relations for the Hamiltonian component fields which of course are to be deduced from the abstract integrability condition (\ref{eq:IntCond}) and conservation equation (\ref{eq:ConEq}). However, with the present reparametrization one will readily transcribe the field equations to the new variables, i.e. the single-particle fields \{${\mathbb{K}}_{a \mu}$, ${\mathbb{L}}_{a \mu}$\} and exchange fields \{$S_{\mu}$, $T_{\mu}$, $F_{\mu}$, $G_{\mu}$\}. In this way there emerges a kind of {\it exchange dynamics} for the latter fields and a {\it single-particle dynamics} for the first fields.
\newline
\newline
\centerline{\it Single-Particle Dynamics}\nopagebreak\nopagebreak

In fact, the integrability condition yields first for the original localization fields $L_{a \mu}$ (\ref{e4:9})
\begin{equation}
\label{e4:39}
\nabla_{\mu}L_{1 \nu}-\nabla_{\nu}L_{1 \mu}= -[\nabla_{\mu}L_{2 \nu}-\nabla_{\nu}L_{2 \mu}]=2 \lambda\, [S_{\mu}T_{\nu}-S_{\nu}T_{\mu}-F_{\mu}G_{\nu}+F_{\nu}G_{\mu}]\;.
\end{equation}
This once more verifies the previous claim (\ref{eq:trL})-(\ref{eq:Lscal}) that the total localization field $L_{\mu}$
\begin{equation}
\label{e4:40}
L_{\mu}\doteqdot L_{1 \mu}+L_{2 \mu}
\end{equation}
actually is a gradient field, in contrast to the individual fields $L_{a \mu}$ ($a=1,2$). Though this result forbids us to conceive the latter fields $L_{a\mu}$ as gradient fields, we nevertheless can introduce individual amplitude fields ${\mathbb{L}}_{a}(x)$ ($a=1,2$) for the effective localization fields ${\mathbb{L}}_{a \mu}$ (\ref{e4:34}) through putting
\begin{equation}
\label{e4:41}
{\mathbb{L}}_{a \mu}=\frac{\partial_{\mu}{\mathbb{L}}_{a}}{{\mathbb{L}}_{a}} \; .
\end{equation}
For the verification of this claim one merely recalls the density dynamics (\ref{e4:11}) and uses therein the new exchange fields  \{$S_{\mu}$, $T_{\mu}$\} in place of the old ones \{$\perpQ_{\mu}$, $\paraN_{\mu}$\}, cf. (\ref{e4:22}). In this way one finds for the single-particle densities $\rho_a$ ($a=1,2$)
\begin{subequations}
\label{e4:42} 
\begin{align}
\partial_{\mu}\,\rho_1 &=2\rho_1 {\mathbb{L}}_{1 \mu} \\
\partial_{\mu}\,\rho_2 &=2\rho_2 {\mathbb{L}}_{2 \mu} 
\end{align}
\end{subequations}
which actually identifies the effective localization fields ${\mathbb{L}}_{a \mu}$ as gradient fields
\begin{equation}
\label{e4:43}
\nabla_{\mu}{\mathbb{L}}_{a \nu}-\nabla_{\nu}{\mathbb{L}}_{a \mu}=0
\end{equation}
in agreement with the claim (\ref{e4:41}).

Once the transition from the localization fields $L_{a \mu}$ to the amplitude fields ${\mathbb{L}}_{a}$ (\ref{e4:41}) has been made, one will not be satisfied with transcribing the abstract conservation equation (\ref{eq:ConEq}) to the localization field $L_{a \mu}$ (or ${\mathbb{L}}_{a \mu}$, resp.) but one will readily proceed to the corresponding wave equations for the scalar amplitudes ${\mathbb{L}}_{a}$ ({\it amplitude equations})
\begin{subequations}
\label{e4:44} 
\begin{align}
\Box \, {\mathbb{L}}_1 &+{\mathbb{L}}_1 \Big\{ \Big( \frac{Mc}{\hbar} \Big)^2-{\mathbb{K}}_{1 \mu}{\mathbb{K}}_1{}^{\mu} \Big\} \nonumber\\
&= \lambda(1-\lambda){\mathbb{L}}_1\Big\{ (S^{\mu}-T^{\mu})(S_{\mu}-T_{\mu})+(F^{\mu}-G^{\mu})(F_{\mu}-G_{\mu}) \Big\} \\[3 mm]
\Box \, {\mathbb{L}}_2 &+{\mathbb{L}}_2 \Big\{ \Big( \frac{Mc}{\hbar} \Big)^2-{\mathbb{K}}_{2 \mu}{\mathbb{K}}_2{}^{\mu} \Big\} \nonumber\\
&= \lambda(1-\lambda) {\mathbb{L}}_2 \Big\{ (S^{\mu}+T^{\mu})(S_{\mu}+T_{\mu})+(F^{\mu}+G^{\mu})(F_{\mu}+G_{\mu}) \Big\} \; .
\end{align}
\end{subequations}
Observe here that these single-particle amplitude equations degenerate to their trivial forms ($a=1,2$)
\begin{equation}
\label{e4:45}
\Box \, {\mathbb{L}}_a +{\mathbb{L}}_a \Big\{ \Big( \frac{Mc}{\hbar} \Big)^2-{\mathbb{K}}_{a \mu}{\mathbb{K}}_a{}^{\mu} \Big\}=0 
\end{equation}
when {\it either} one deals with pure states ($\lambda \equiv 1$) {\it or} when one neglects the exchange fields ($S_{\mu}=T_{\mu}=F_{\mu}=G_{\mu}\equiv0$). For such a situation, the degenerate equations (\ref{e4:45}) can be shown to be equivalent to two decoupled Klein-Gordon equations \cite{RuHuSo} for two single-particle wave functions $\psi_a(x)$ ($a=1,2$):
\begin{equation}
\label{e4:46}
D^{\mu}D_{\mu}\psi_a+\Big(\frac{Mc}{\hbar}\Big)^2\psi_a=0
\end{equation}
\centerline{($D_{\mu}\psi_a\doteqdot\partial_{\mu}\psi_a-iA_{a \mu}\psi_a$).} 
Evidently, the two particles are now disentangled and thus can interact exclusively via the ordinary gauge forces mediated by the corresponding gauge potentials $A_{a \mu}$. Either of the two wave functions $\psi_a$ is then of the usual form
\begin{subequations}
\label{e4:47} 
\begin{align}
\psi_1 &= {\mathbb{L}}_1 e^{-i\alpha_1}\\
\psi_2 &= {\mathbb{L}}_2 e^{-i\alpha_2}
\end{align}
\end{subequations}
where the phases $\alpha_a$ are given by
\begin{equation}
\label{e4:48}
\alpha_a=\int^xdx^{\mu}({\mathbb{K}}_{a \mu}-A_{a \mu}) \;.
\end{equation} 

But clearly the typical quantum effects are expected to occur in connection with the simultaneous presence of non-trivial exchange fields and mixtures ($\lambda \not=1$). Here the two independent Klein-Gordon equations (\ref{e4:45})-(\ref{e4:46}) become modified to their ``entangled'' forms (\ref{e4:44}) yielding a coupling of the single-particle amplitudes ${\mathbb{L}}_a$ to the exchange fields with the occurrence of certain {\it exchange effects} (separate paper).

A similar coupling of single-particle and exchange fields is naturally to be expected also for the kinetic fields. Indeed, the following curl relations for the effective kinetic fields ${\mathbb{K}}_{a \mu}$ ($a=1,2$) are deduced from the abstract integrability condition (\ref{eq:IntCond}) in a straightforward manner: 
\begin{subequations}
\label{e4:49} 
\begin{align}
\nabla_{\mu}{\mathbb{K}}_{1 \nu}-\nabla_{\nu}{\mathbb{K}}_{1 \mu}&= F_{1 \mu \nu}+2\lambda (1-\lambda)\,\Big[ (S_{\mu}-T_{\mu})(F_{\nu}-G_{\nu})-(S_{\nu}-T_{\nu})(F_{\mu}-G_{\mu}) \Big]\\
\nabla_{\mu}{\mathbb{K}}_{2 \nu}-\nabla_{\nu}{\mathbb{K}}_{2 \mu}&= F_{2 \mu \nu}+2\lambda (1-\lambda)\,\Big[ (S_{\mu}+T_{\mu})(F_{\nu}+G_{\nu})-(S_{\nu}+T_{\nu})(F_{\mu}+G_{\mu}) \Big] \;.
\end{align}
\end{subequations}
From here it becomes obvious that the effective kinetic fields ${\mathbb{K}}_{a \mu}$ do not only ``feel'' the corresponding components $F_{ a \mu \nu}$ of the bundle curvature $\F_{\mu \nu}$ (\ref{e4:4}a) but they feel an additional {\it exchange field strength}, namely the second terms on the right of equation (\ref{e4:49}) which is built up essentially by the exchange fields $S_{\mu}$, $T_{\mu}$, $F_{\mu}$, $G_{\mu}$. However for the pure states ($\lambda=1$) this exchange effect vanishes again and thus we arrive at the same result as mentioned previously in connection with the energy-momentum density $T_{\mu \nu}$ (\ref{e4:35}) and the amplitude equations (\ref{e4:44}), namely that the exchange effects can become active only for the mixtures but not for the pure states. Observe however that the present exchange effect, occuring for the effective kinetic fields ${\mathbb{K}}_{a \mu}$ (\ref{e4:49}), is somewhat different in comparison to the preceding situations of energy-momentum density ${}^{(a)}t_{\mu \nu}$ (\ref{e4:38}) and amplitude equation (\ref{e4:44}); the reason for this is that the kinetic exchange term (\ref{e4:49}) vanishes if one of the two exchange pairs \{$S_{\mu}$, $T_{\mu}$\} or \{$F_{\mu}$, $G_{\mu}$\} vanishes whereas the other two exchange effects can survive also with only one of the two exchange pairs being present.
 
In order to close the single-particle dynamics for the effective fields, one has to write down the source relations for the kinetic fields. The desired relations are obtained again in a straightforward way from the abstract conservation equation (\ref{eq:ConEq}):
\begin{subequations}
\label{e4:50} 
\begin{align}
\nabla^{\mu}{\mathbb{K}}_{1 \mu}+2{\mathbb{L}}_{1}{}^{\mu}\, {\mathbb{K}}_{1 \mu} &=0 \\
\nabla^{\mu}{\mathbb{K}}_{2 \mu}+2{\mathbb{L}}_{2}{}^{\mu}\, {\mathbb{K}}_{2 \mu} &=0 \; .
\end{align}
\end{subequations} 
Observing here the gradient property of the effective localization fields ${\mathbb{L}}_{a \mu}$ (\ref{e4:41}), one easily recognizes the present source relations (\ref{e4:50}) as nothing else than the former charge conservation laws (\ref{eq:chcon}) when the result (\ref{e4:32}) for the currents $j_{a \mu}$ is observed together with the density derivatives (\ref{e4:42}).
\newline
\newline
\centerline{\it Exchange Dynamics}\nopagebreak

As the preceding single-particle dynamics demonstrates, there is a coupling of the single-particle fields to the exchange fields, see the amplitude equations (\ref{e4:44}) and curl relations (\ref{e4:49}) for the effective kinetic fields. This coupling shows that the single-particle dynamics decouples from the exchange fields only for the pure states ($\lambda=1$) but not for the mixtures ($\lambda \not=1$). Thus, when dealing with the pure states, one can first solve the autonomous single-particle system and substitute this as a kind of rigid background into the exchange dynamics \cite{RuHuSo}. However for the mixtures, one must solve the single-particle system in one step together with the exchange system because also the latter system couples back to the first system. This coupling is explicitly seen by writing down the source relations for the exchange fields (deducible again from the conservation equation (\ref{eq:ConEq})):
\nopagebreak
\begin{subequations}
\label{e4:51} 
\begin{align}
\nabla^{\mu}S_{\mu} =&-(S^{\mu}-T^{\mu}){\mathbb{L}}_{1 \mu}-(S^{\mu}+T^{\mu}){\mathbb{L}}_{2 \mu}+(F^{\mu}-G^{\mu}){\mathbb{K}}_{1 \mu}+(F^{\mu}+G^{\mu}){\mathbb{K}}_{2 \mu} \nonumber\\
& +2(2\lambda-1)S^{\mu}S_{\mu}-2\lambda F^{\mu}F_{\mu}-2(\lambda-1)G^{\mu}G_{\mu}\\[3 mm]
\nabla^{\mu}T_{\mu} =&(S^{\mu}-T^{\mu}){\mathbb{L}}_{1 \mu}-(S^{\mu}+T^{\mu}){\mathbb{L}}_{2 \mu}-(F^{\mu}-G^{\mu}){\mathbb{K}}_{1 \mu}+(F^{\mu}+G^{\mu}){\mathbb{K}}_{2 \mu} \nonumber\\
& +2(2\lambda-1)(S^{\mu}T_{\mu}-F^{\mu}G_{\mu})\\[3 mm]
\nabla^{\mu}F_{\mu} =&-(F^{\mu}-G^{\mu}){\mathbb{L}}_{1 \mu}-(F^{\mu}+G^{\mu}){\mathbb{L}}_{2 \mu}-(S^{\mu}-T^{\mu}){\mathbb{K}}_{1 \mu}-(S^{\mu}+T^{\mu}){\mathbb{K}}_{2 \mu} \nonumber\\
& +2(\lambda-1)G^{\mu}T_{\mu}+2(3\lambda-1)F^{\mu}S_{\mu}\\[3 mm]
\nabla^{\mu}G_{\mu} =&(F^{\mu}-G^{\mu}){\mathbb{L}}_{1 \mu}-(F^{\mu}+G^{\mu}){\mathbb{L}}_{2 \mu}+(S^{\mu}-T^{\mu}){\mathbb{K}}_{1 \mu}-(S^{\mu}+T^{\mu}){\mathbb{K}}_{2 \mu} \nonumber\\
& +2(3\lambda-2)S^{\mu}G_{\mu}+2\lambda F^{\mu}T_{\mu} \;. 
\end{align}
\end{subequations}
Observe here that even for the pure states ($\lambda=1$) the exchange fields do not decouple from the single-particle fields. Such a decoupling effect occurs exclusively for the curl relations (to be deduced from the integrability condition(\ref{eq:IntCond})):
\begin{subequations}
\label{e4:52} 
\begin{align}
\nabla_{\mu}S_{\nu}-\nabla_{\nu}S_{\mu}&=0 \\
\nabla_{\mu}T_{\nu}-\nabla_{\nu}T_{\mu}&=2(2\lambda-1)[S_{\mu}T_{\nu}-S_{\nu}T_{\mu}]+2[G_{\mu}F_{\nu}-G_{\nu}F_{\mu}]\\
\nabla_{\mu}F_{\nu}-\nabla_{\nu}F_{\mu}&=2(\lambda-1)[S_{\mu}F_{\nu}-S_{\nu}F_{\mu}+G_{\mu}T_{\nu}-G_{\nu}T_{\mu}]\\
\nabla_{\mu}G_{\nu}-\nabla_{\nu}G_{\mu}&=2\lambda [F_{\mu}T_{\nu}-F_{\nu}T_{\mu}+S_{\mu}G_{\nu}-S_{\nu}G_{\mu}] \; .
\end{align}
\end{subequations}
Here, the first equation (\ref{e4:52}a) says that $S_{\mu}$ must always be a gradient field which however is clear from its very definition (\ref{e4:21}). Furthermore the vector field $F_{\mu}$ becomes also a gradient field for the pure states ($\lambda=1$), cf. (\ref{e4:52}c).

Thus the new parametrization yields not only a formal simplification of the whole RST dynamics but makes also the theory more transparent by providing us with a clear subdivision of the field variables into single-particle and exchange objects. As a consequence, one becomes able to discuss the RST predictions in terms of single-particle concepts and exchange interactions.

\newpage
\section{Bundle Geometry}

Whenever a superselection rule is seen to be in action, such as the present dynamical separation of positive and negative mixtures, one tends to guess that such a strict rule might have to do something with the global topological features of the corresponding field configurations. Indeed it is easily seen from fig.1 that the one-parted hyperboloid $H_{(-)}$, being infinitely connected, cannot be continuously deformed into the two-parted hyperboloid $H_{(+)}$ whose individual parts are simply connected. Therefore it is self-suggesting to guess that this topological arrangement is somehow encoded into the field variables of positive and negative mixtures.  Thus there arises an attractive mathematical problem, namely to identify the place where the two different topologies are buried in the space-time dependence of the field variables.
 
 The answer to this question is the following: Take the renormalization factor $Z_O(x)$ (\ref{e4:18}c) and the mixture variable $\lambda(x)$ (\ref{e4:24}) together with the 1-forms ${\mathbf F}=\{F_{\mu}\}$ and $\mathbf{G}=\{G_{\mu}\}$, cf. (\ref{e4:30}), and construct the 2-form ${\mathbf E}=\{E_{\mu \nu}\}$
\begin{equation}
\label{e5:1}
{\mathbf E}=4 \frac{\lambda}{Z_O}{\mathbf G}\wedge {\mathbf F} \; ,
\end{equation}
i.e. in components
\begin{equation}
\label{e5:2}
E_{\mu \nu}=4 \frac{\lambda}{Z_O}[G_{\mu}F_{\nu}-G_{\nu}F_{\mu}] \; .
\end{equation}
Then use the field equations for the mixture variable $\lambda$ (\ref{e4:25}) and renormalization factor $Z_O$ (\ref{e4:23}c) together with the curl relations for ${\mathbf F}$ and ${\mathbf G}$ (\ref{e4:52}) in order to show that the 2-form ${\mathbf E}$ is closed: ${\mathbf dE}=0$, i.e. in components
\begin{equation}
\label{e5:3}
\nabla_{\lambda}E_{\mu \nu}+\nabla_{\mu}E_{\nu \lambda}+\nabla_{\nu}E_{\lambda \mu}=0 \; .
\end{equation}
Actually ${\mathbf E}$ is an element of the deRham cohomology group $H^{2}(M_4, {\mathbb{Z}})$ over space-time ($M_4$), and its value upon some 2-cycle $C^2$ is a fixed integer $z[C^2]$
\begin{equation}
\label{e5:4}
z[C^2]=\oint_{C^2}{\mathbf E}
\end{equation}
being independent of continuous deformations of the closed 2-surface $C^2$. Furthermore, for the present 2-particle case ($N=2$) the ``winding numbers'' $z[C^2]$ are found to be always zero for both types of mixtures, no matter which element $C^2$ of the corresponding homology group of the underlying space-time is taken as the integral surface. Thus ${\mathbf E}$ turns out to be an exact 2-form: ${\mathbf E}={\mathbf d}{\mathbf e}$, in components
\begin{equation}
\label{e5:5}
E_{\mu \nu}=\nabla_{\mu}e_{\nu}-\nabla_{\nu}e_{\mu} \; ,
\end{equation}
with ${\mathbf e}=\{e_{\mu}\}$ some appropriate (non-unique) 1-form.

Though both mixtures have the same (trivial) winding numbers $z[C^2]$, the 2-form ${\mathbf E}$ carries nevertheless different topological information. It can namely be shown that ${\mathbf E}$ is the {\it Euler class} of appropriate fibre bundles ({\it density bundles}, to be constructed below). This means concretely that there exists a regular ``density map'' (${\thetavec}$) from space-time to the hyperboloids $H_{(\pm)}$ in 3-dimensional, pseudo-Euclidian density configuration space (fig.1)
\begin{subequations}
\label{e5:6}
\begin{align}
\thetavec :\quad \quad \theta^j&=\theta^j(x)\; ,\quad (j=1,2,3)\\
\theta^j \theta_j&=\sigma_{\ast}
\end{align}
\end{subequations}
such that the 2-form ${\mathbf E}$ (\ref{e5:1}) yields just the 2-volume form on those hyperboloids (up to an exact 2-form ${\mathbf d}{}^{(||)}{\mathbf B}$):
\begin{equation}
\label{e5:7}
\sigma{\ast}\cdot E_{\mu \nu}=\nabla_{\mu}\paraB_{\nu}-\nabla_{\nu}\paraB_{\mu}+\theta_j\epsilon^j{}_{kl}(\partial_{\mu}\theta^k)(\partial_{\nu}\theta^l) \; .
\end{equation}
Therefore the winding numbers (\ref{e5:4}) measure the surface content (on the hyperboloids $H_{(\pm)}$) of the image of the 2-cycle $C^2$ with respect to the density map $\thetavec$ (\ref{e5:6}). But since both hyperboloids $H_{(\pm)}$ as non-compact 2-surfaces have trivial second homotopy group ($\Pi_2(H_{(\pm)})=0$), the image of the compact $C^2$ (e.g. 2-sphere $S^2$) under $\thetavec$ yields a multiple covering (with alternative orientations) of some subspace of $H_{(\pm)}$ and this must necessarily imply the vanishing of the winding numbers $z[C^2]$ (\ref{e5:4}). Would the density map $\thetavec$ lead into some image space with non-trivial second homotopy group $\Pi_2$ (, as it may be expected for particle number $N>2$), the winding numbers of positive and negative mixtures could well be different in a non-trivial way.

In what follows a proof is presented for all these claims, leading directly to the central result (\ref{e5:7}).

\subsection{Density Bundle}
For a deeper understanding of the mixture topology it is highly instructive to evoke the mathematical tool of fibre bundles. Within this framework, one recovers not only the decisive curl relation (\ref{e5:3}) (which one could find also by merely applying the dynamical equations in a straight-forward manner) but one becomes capable to identify that relation as the {\it Bianchi identity} for the bundle curvature ${\mathbf E}$. In this way one arrives at the true origin of the winding numbers $z[C^2]$ (\ref{e5:4}).

For the construction of the fibre bundle in question, recall first that the intensity matrix $\I$, as an Hermitian $2 \times 2$-matrix, properly speaking has four independent components. These however could be reduced to only three real components by resorting to a co-moving reference system (the {\it RTB basis} \cite{MaRuSo}), so that the intensity matrix $\I$ reappears as a 3-component object, cf. (\ref{e4:5}). As the three remaining components one may take the overlap density $s$ (\ref{e4:7}), the total density $\rho$ (\ref{e4:12}) and the relative density $q$ (\ref{e4:13}) which obey the density dynamcis (\ref{e4:11}) and give rise to the introduction of the renormalization factors $Z_{T}$, $Z_{R}$, $Z_O$ (\ref{e4:18}). The interesting point here is now that the latter three objects may be collected into a 3-vector $\thetavec=\{\theta^j\}$ ($j=1,2,3$) as an element of a 3-dimensional real vector fibre over any event $x$ of space-time: 
\begin{subequations}
\label{e5:8} 
\begin{align}
\theta^1 & \doteqdot Z_T\\
\theta^2 & \doteqdot Z_R\\
\theta^3 & \doteqdot Z_O \; .
\end{align}
\end{subequations}
This 3-dimensional vector fibre is equipped with a Minkowskian fibre metric $\eta=diag(1,-1,-1)$ so that the normalization constraint (\ref{e5:6}b) simply says that $\thetavec$ (\ref{e5:8}) is a unit vector
\begin{equation}
\label{e5:9}
\eta_{jk}\theta^j \theta^k \equiv \theta_j \theta^j=\sigma_{\ast} \; ,
\end{equation}
namely either ``time-like'' ($\sigma_{\ast}=+1$) or ``space-like'' ($\sigma{\ast}=-1$), resp. ``light-like'' ($\sigma_{\ast}=0$). 

The next interesting point comes into play when one wishes to equip this density bundle ($\thetavec$) with a connection ($\B_{\mu}$, say). In order to preserve the normalization (\ref{e5:9}), the connection 1-form $\B_{\mu}$ must take its values in the 3-dimensional Lorentz algebra $\GO(1,2)$. The latter algebra may be thought to be spanned by three anti-Hermitian generators $\B_j=-{\bar{\mathcal B}}_j$ where the concept of Hermiticity refers here to the Minkowskian metric $\eta$. This means that when $B^k{}_{lj}$ are the elements of the matrix $\B_j$ and $\bar{B}^k{}_{lj}$ are the elements of $\bar{\B}_j$ then we have 
\begin{equation}
\label{e5:10}
\bar{B}^k{}_{lj}=\eta_{lm}B^m{}_{nj}\eta^{nk} \;.
\end{equation}
The standard solution of this requirement for the generators is the well-known permutation symbol $\epsilonvec$ as the completely antisymmetric unit tensor object of the third rank
\begin{equation}
\label{e5:11}
B^k{}_{lj}=-\epsilon^k{}_{lj} \equiv-\eta^{km}\epsilon_{mlj} \;.
\end{equation}
The corresponding commutation relations are easily found as
\begin{equation}
\label{e5:12}
[\B_j,\B_k]=\epsilon^l{}_{jk}\B_l \; .
\end{equation}
Thus the desired bundle connection $\B_{\mu}$ of ($\thetavec$), being defined through
\begin{equation}
\label{e5:13}
\partial_{\mu}\thetavec=-\B_{\mu}\cdot \thetavec
\end{equation}
\nopagebreak\centerline{(in components: $\partial_{\mu}\theta^j=-B^j{}_{k \mu}\theta^k$),}
may be decomposed with respect to the anti-Hermitian basis $\{\B_j\}$ as 
\begin{equation}
\label{e5:14}
\B_{\mu}=B^j{}_{\mu}\B_j
\end{equation}
\nopagebreak\centerline{(in components: $B^j{}_{k \mu}=-B^l{}_{\mu}\epsilon^j{}_{k l}$).}

In order to explicitly determine the connection $\B_{\mu}$, one could think now that the connection coefficients $B^j{}_{\mu}$ could simply be read off from the renormalization dynamics (\ref{e4:23}) which would yield the following result
\begin{subequations}
\label{e5:15} 
\begin{align}
B^1{}_{\mu}&=2 \perpQ_{\mu} \\
B^2{}_{\mu}&=-2 \paraN_{\mu} \\
B^3{}_{\mu}&=l_{\mu} \; . 
\end{align}
\end{subequations}
However the problem is here that this connection is not unique since one can add to any $\B_{\mu}$ some $\GO(1,2)$ element ($\Z$, say) which annihilates the 3-vector $\thetavec$, i.e. $\Z\cdot\thetavec=0$:
\begin{equation}
\label{e5:16}
\B_{\mu} \Rightarrow \, '\B_{\mu}=\B_{\mu} - b_{\mu}\cdot \Z \; .
\end{equation}
Here, the ${\mathbb{R}}^1$-valued 1-form $b_{\mu}$ is arbitrary and $\Z$ is easily seen to decompose with respect to the standard basis (\ref{e5:11}) as follows
\begin{equation}
\label{e5:17}
\Z=\theta^j \B_j \; .
\end{equation}

\subsection{Gauges and Projections}
This non-uniqueness of the connection $\B_{\mu}$ in the density bundle ($\thetavec$) must not necessarily be considered as a deficiency of the theory but may rather be interpreted as the presence of a gauge degree of freedom inherent in the bundle formalism. This means that we admit $SO(1,2)$-valued gauge transformations $\S(x)$ which act homogeneously over the vector fibre in the following way
\begin{subequations}
\label{e5:18} 
\begin{align}
\thetavec &\Rightarrow \thetavec ' =\S \cdot \thetavec \\
D_{\mu}\, \thetavec &\Rightarrow D_{\mu}\, \thetavec '= \S \cdot D_{\mu} \, \thetavec
\end{align}
\end{subequations}
\centerline{(i.e. in components $\theta '{}^j=S^j{}_k \theta^k$).}
These gauge transformations induce an inhomogeneous transformation law for the bundle connection $\B_{\mu}$:
\begin{equation}
\label{e5:19}
\B_{\mu} \Rightarrow \B_{\mu}'=\S \cdot \B_{\mu} \cdot \S^{-1} + \S \cdot \partial_{\mu} \, \S^{-1} \; .
\end{equation}\

Now we try to restrict the gauge group to a subgroup of $SO(1,2)$ in such a way that the original non-uniqueness relation (\ref{e5:16}) becomes identical to the present gauge transformation (\ref{e5:19}). This requirement fixes the elements $\S$ of the restricted gauge group as
\begin{equation}
\label{e5:20}
\S=exp[b{(x)}\Z] \; ,
\end{equation}
i.e. the abelian rotations in the orthogonal 2-space of the density vector $\thetavec$ which itself is invariant under the restricted gauge group:
\begin{equation}
\label{e5:21}
\S \cdot \thetavec = \thetavec \; .
\end{equation}
Furthermore, it can be shown by some simple bundle analysis that the 1-form $b_{\mu}$, emerging in the non-uniqueness relation (\ref{e5:16}), is changed by the derivative of the group parameter $b$ (\ref{e5:20}) into $b_{\mu}'$:
\begin{equation}
\label{e5:22}
b_{\mu} \Rightarrow b_{\mu}'=b_{\mu}+ \partial_{\mu}\, b \; .
\end{equation}
This analysis becomes quite easy by resorting to the fact that the generator of rotations $\Z$ (\ref{e5:17}) is covariantly constant
\begin{equation}
\label{e5:23}
\D_{\mu} \Z = 0
\end{equation}
\centerline{($\D_{\mu} \Z \doteqdot \partial_{\mu} \, \Z + [\B_{\mu},\Z]$),}
or in components:
\begin{equation}
\label{e5:24}
D_{\mu} \theta^j \equiv 0
\end{equation}
\centerline{($D_{\mu}\theta^j \doteqdot \partial_{\mu} \, \theta^j +\epsilon^j{}_{kl}B^k{}_{\mu} \theta^l$).}

It is important to remark here that the restricted gauge transformations (\ref{e5:20}) change the 1-form $b_{\mu}$ in the non-uniqueness relation (\ref{e5:16}) by a gradient field $\partial_{\mu}\, b$, cf. (\ref{e5:22}), and therefore $b_{\mu}$ can be gauged off exclusively when the original $b_{\mu}$ is also a gradient field. On the other hand, one can take advantage of the non-uniqueness of $\B_{\mu}$ by letting vanish its component $\paraB_{\mu}$ relative to the gauge algebra \{$\Z$\}
\begin{equation}
\label{e5:25}
\paraB_{\mu}=-\frac{1}{2} \, tr(\Z\cdot\B_{\mu}) \; .
\end{equation} 
In this way one is lead to an other connection (the {\it projective connection} $\P_{\mu}$) which is obtained from  the original $\B_{\mu}$ by subtracting off its non-unique $\Z$ part, i.e. one puts
\begin{equation}
\label{e5:26}
\P_{\mu} =\B_{\mu} + \frac{1}{2} \, \sigma_{\ast} \Z \cdot tr(\Z\cdot \B_{\mu})=\B_{\mu}-\sigma_{\ast}\paraB_{\mu}\cdot \Z \; .
\end{equation}
This does not change the topological properties of the bundle connection because the Euler class $\mathbf E$ (\ref{e5:7}) is modified merely by some exact 2-form ${\mathbf d}{\,{}^{(||)}{\mathbf B}}$ and this of course cannot change the winding numbers $z[C^2]$ (\ref{e5:4}). The reason is that the integral of an exact 2-form over some cycle $C^2$ always must vanish
\begin{equation}
\label{e5:27}
\oint_{C^2}{\mathbf d}{\,{}^{(||)}{\mathbf B}} \equiv 0 \; .
\end{equation}

Since this projective connection $\P_{\mu}$ (\ref{e5:26}) plays an important part for the subsequent method of bundle trivialization, it is convenient to display some of its geometric properties. First the projector $\P=\{P^j{}_k\}$ is defined by the usual relations
\begin{subequations}
\label{e5:28} 
\begin{align}
\P \cdot \thetavec &=0 \\
\P \cdot \P &=\P=\bar{\P} \\
tr \P &=2 \; ,
\end{align}
\end{subequations}
and furthermore it is related to the square of the rotation generator $\Z$ (\ref{e5:17}) through 
\begin{equation}
\label{e5:29}
\P=-\sigma_{\ast}\cdot \Z^2 \; .
\end{equation}
It can also be expressed in terms of the 3-vector $\thetavec$ as
\begin{equation}
\label{e5:30}
\P={\mathbf 1}-\sigma_{\ast}\, \thetavec \otimes \bar{\thetavec} \; ,
\end{equation}
and finally the corresponding projective connection $\P_{\mu}$ in the density bundle ($\thetavec$) can then be written down in terms of the projector $\P$ as 
\begin{equation}
\label{e5:31}
\P_{\mu}=[\P,\partial_{\mu} \, \P]=[\P,[\P,\B_{\mu}]]
\end{equation}
from which it is immediately seen that $\P_{\mu}$ has vanishing component relative to the gauge algebra spanned by the generator $\Z$:
\begin{equation}
\label{e5:32}
tr (\Z\cdot \P_{\mu})=0 \; .
\end{equation}

The projector $\P$ is also helpful for considering the nature of the restricted gauge group generated by the elements $\S$ (\ref{e5:20}) which read in terms of $\P$ and $\Z$
\begin{equation}
\label{e5:33}
\S= \left\{\begin{array}{l}
1-\P+\cos b \cdot \P + \sin b\cdot \Z \; \mbox{, } \quad \sigma_{\ast}=+1\\ 
1-\P+\cosh b \cdot \P + \sinh b\cdot \Z \; \mbox{, }\quad \sigma_{\ast}=-1 \; .
\end{array}\right.
\end{equation}
This result says that the restricted gauge group for the positive mixtures ($\sigma_{\ast}=+1$) is the ordinary rotation group $SO(2)$ and for the negative mixtures ($\sigma_{\ast}=-1$) it is the Lorentz group  $SO(1,1)$ of ($1+1$) dimensional Minkowskian space. But apart from this difference with respect to the gauge group, the projective bundle geometry is the same for both kinds of mixtures. 

\subsection{Curvature}
Though the  preceding projection mechanism for constructing the density bundle ($\thetavec$) may tempt one to think that both kinds of mixtures share the same geometric structures, the subsequent discussion of the bundle curvature will clearly reveal some essential differences.

However in the first instance, the bundle curvature ($\C_{\mu \nu}$, say) looks still identical for both situations. The general definition of the curvature operator is
\begin{equation}
\label{e5:34}
\C_{\mu \nu}=\nabla_{\mu}\B_{\nu}-\nabla_{\nu}\B_{\mu}+[\B_{\mu},\B_{\nu}]
\end{equation}
and this operator measures the non-commutativity of the gauge covariant derivatives, e.g. in form of the bundle identity for the generator $\Z$ (\ref{e5:17})
\begin{equation}
\label{e5:35}
[\D_{\mu}\D_{\nu}-\D_{\nu}\D_{\mu}]\Z \equiv [\C_{\mu \nu},\Z] \; .
\end{equation} 
Now according to the well-kown Ambrose-Singer theorem \cite{KoNo}, the curvature operator $\C_{\mu \nu}$ spans the algebra of the holonomy group which in general is a subgroup of the gauge group. Since, for the present situation, the latter group does not have non-trivial subgroups, the curvature operator must take its values in the gauge algebra itself, i.e. in the rotation algebra $\GO(2)$ for the positive mixtures ($\sigma_{\ast}=+1$) and in the Lorentz algebra $\GO(1,1)$ for the negative mixtures ($\sigma_{\ast}=-1$), see equation (\ref{e5:33}). This however implies that the curvature is proportional to that generator (i.e. $\C_{\mu \nu} \sim \Z$); and indeed if one substitutes the present bundle connection $\B_{\mu}$ (\ref{e5:15}) into the general curvature definition (\ref{e5:34}), a straightforward calculation yields with the help of the curl relations (\ref{e4:39}) and (\ref{e4:52}) the following result involving the Euler class $E_{\mu \nu}$ (\ref{e5:2})
\begin{equation}
\label{e5:36}
\C_{\mu \nu}=E_{\mu \nu} \cdot \Z \; .
\end{equation}
Clearly this result makes the bundle identity (\ref{e5:35}) consistent with the covariant constancy of the generator $\Z$ (\ref{e5:23}).

Furthermore, the general structure of curvature $\C_{\mu \nu}$ (\ref{e5:34}) implies the well-known Bianchi identity
\begin{equation}
\label{e5:37}
\D_{\lambda}\C_{\mu \nu}+\D_{\mu}\C_{\nu \lambda}+\D_{\nu}\C_{\lambda \mu} \equiv 0
\end{equation} 
which is immediately transcribed to the curvature coefficient $E_{\mu \nu}$ as shown by the former equation (\ref{e5:3}). In this way one has revealed the true origin why that 2-form $E_{\mu \nu}$ must be closed. This closedness property must appear as a mere incident when one can resort exclusively to the dynamical equations.

The curvature result (\ref{e5:36}) says that the curvature 2-form $\C_{\mu \nu}$ takes its values in the 1-dimensional algebra \{$\Z$\} of (pseudo-)rotations and thus sweeps out a more restricted range than its connection $\B_{\mu}$. The latter takes its values in the Lorentz-algebra $\GO (1,2)$ and therefore one can on principle apply $SO(1,2)$-valued gauge transformations. The subgroup of gauge transformations $\S$ (\ref{e5:20}), applied so far, thus turns out as the {\it holonomy group} of the density bundle ($\thetavec$).

\subsection{Bundle Trivialization}
So far, there evidently appears no difference in the geometric structure of both types of mixtures, but actually the curvature coefficient $E_{\mu \nu}$ carries different topological characteristics for either case. In order to see this more clearly, one identifies that curvature coefficient $E_{\mu \nu}$ as the pullback (with respect to the density map $\thetavec$ (\ref{e5:6})) of the volume cell on the hypersurface $H_{(\pm)}$ in density configuration space (fig.1). The desired identification is obtained by embedding the non-trivial density bundle ($\thetavec$) into some higher-dimensional {\it trivial} bundle such that ($\thetavec$) reappears as a reduced bundle \cite{KoNo}.

According to this line of arguments, one adds to the non-trivial connection $\B_{\mu}$ in the density bundle ($\thetavec$) some element $\E_{\mu}$ of the embedding Lie algebra such that the resulting connection ($\Bcirc_{\mu}$, say) is trivial. More concretely, we put 
\begin{equation}
\label{e5:38}
\Bcirc_{\mu} = \B_{\mu} +\E_{\mu}
\end{equation} 
where $\E_{\mu}$ must transform {\it homogeneously}
\begin{equation}
\label{e5:39}
\E_{\mu} \Rightarrow \E'_{\mu} = \S \cdot \E_{\mu} \cdot \S^{-1}
\end{equation} 
in order that the new connection $\Bcirc_{\mu}$ can transform {\it inhomogeneously} as is generally required for bundle connections (see (\ref{e5:19}))
\begin{equation}
\label{e5:40}
\Bcirc_{\mu} \Rightarrow \Bcirc'_{\mu} = \S \cdot \Bcirc_{\mu} \cdot \S^{-1} +\S \cdot \partial_{\mu} \S^{-1} \; .
\end{equation} 
The point with this change of connection is here that the new curvature ($\Ccirc_{\mu \nu}$, say) 
\begin{equation}
\label{e5:41}
\Ccirc_{\mu \nu} = \nabla_{\mu} \Bcirc_{\nu} -\nabla_{\nu} \Bcirc_{\mu} +[\Bcirc_{\mu},\Bcirc_{\nu}]
\end{equation}
is required to vanish ($\Ccirc_{\mu \nu} \equiv 0$), i.e.
\begin{equation}
\label{e5:42}
0=\Ccirc_{\mu \nu} = \C_{\mu \nu} + \D_{\mu} \E_{\nu} - \D_{\nu} \E_{\mu} +[\E_{\mu},\E_{\nu}] \; .
\end{equation}
This allows us to recast the original bundle curvature $\C_{\mu \nu}$ into the following form
\begin{equation}
\label{e5:43}
\C_{\mu \nu} = -(\D_{\mu} \E_{\nu} -\D_{\nu} \E_{\mu} +[\E_{\mu},\E_{\nu}])
\end{equation}
\centerline{($\D_{\mu} \E_{\nu} \doteqdot \nabla_{\mu} \E_{\nu} +[\B_{\mu},\E_{\nu}]$).}
This structure is then inherited also by the curvature coefficient $E_{\mu \nu}$ (\ref{e5:36}).

Indeed, the general structure of $E_{\mu \nu}$ follows directly from the latter form (\ref{e5:43}) of the curvature $\C_{\mu \nu}$; this is immediately seen by inverting the former curvature relation (\ref{e5:36}) as
\begin{equation}
\label{e5:44}
E_{\mu \nu}= -\frac{1}{2} \sigma_{\ast} tr(\Z \cdot \C_{\mu \nu})
\end{equation}
and then substituting herein the new form of $\C_{\mu \nu}$ (\ref{e5:43}). This procedure finally yields for the curvature coefficient $E_{\mu \nu}$
\begin{equation}
\label{e5:45}
E_{\mu \nu}=\sigma_{\ast} \cdot [\nabla_{\mu}\paraB_{\nu}-\nabla_{\nu}\paraB_{\mu}]+\frac{1}{2} \sigma_{\ast} tr(\Z \cdot [\E_{\mu},\E_{\nu}])
\end{equation}
with the 1-form ${}^{(||)}{\mathbf B}={}^{(||)}{B}_{\mu} {\mathbf d}x^{\mu}$ being given by (\ref{e5:25}) and expressed here in terms of $\E_{\mu}$
\begin{equation}
\label{e5:46}
\paraB_{\mu}=\frac{1}{2} tr(\Z \cdot \E_{\mu}) \; .
\end{equation}

This latter relation hints at the possibility of projective generation of the non-trivial  density bundle ($\thetavec$) from a trivial embedding bundle, namely through the fact that the component of the trivial connection $\Bcirc_{\mu}$ (\ref{e5:38}) with respect to the holonomy algebra \{$\Z$\} is zero:
\begin{equation}
\label{e5:47}
tr(\Z\cdot \Bcirc_{\mu}) = tr \{\Z \cdot (\B_{\mu}+\E_{\mu}) \} =0 \; ,
\end{equation}
just as was the case with the projective connection $\P_{\mu}$, see equation (\ref{e5:32}). Evidently the present result (\ref{e5:45}) is nothing else than the former claim for $E_{\mu \nu}$ (\ref{e5:7}) provided we can prove the coincidence of both second terms on the right-hand sides of equations (\ref{e5:7}) and (\ref{e5:45}), i.e. it remains to prove the following identity
\begin{equation}
\label{e5:48}
\frac{1}{2} tr(\Z \cdot [\E_{\mu},\E_{\nu}]) \equiv \theta_j \epsilon^j{}_{kl}(\partial_{\mu}\theta^k)(\partial_{\nu} \theta^l)\; .
\end{equation}

Now it should be clear that this final step may well bring into play some differences between the positive and negative mixtures. Indeed, we will readily see that the expected difference consists in the specific way of embedding the density bundle ($\thetavec$) into some {\it trivial} bundle of higher dimension.

\subsection{Euler Class}
Thus the goal is now to find the bundle of smallest fibre dimension which can carry a {\it trivial} but non-vanishing connection $\Bcirc_{\mu}$, whose curvature $\Ccirc_{\mu \nu}$ (\ref{e5:42}) vanishes together with the projection of its connection $\Bcirc_{\mu}$ to the holonomy subalgebra, cf. (\ref{e5:47}).

In order to get some ansatz for the desired $\Bcirc_{\mu}$, one recalls that any trivial connection can be written as a ``pure gauge'', i.e. the pullback of the Maurer-Cartan form over the gauge group:
\begin{equation}
\label{e5:49}
\Bcirc_{\mu} = \Lambda \cdot \partial_{\mu} \Lambda^{-1} \; .
\end{equation}
Here $\Lambda$ represents an element of that Lie group which is generated by the Lie-algebra valued range of $\Bcirc_{\mu}$. Of course the most simple guess for $\Lambda$ would be just the element $\S$ (\ref{e5:20}) of the holonomy group itself ($\Lambda \Rightarrow \S$); in this case one would obtain for $\Bcirc_{\mu}$ (\ref{e5:49}) the following result for {\it positive} mixtures ($\sigma_{\ast}=+1$):
\begin{equation}
\label{e5:50}
\Bcirc_{\mu} \Rightarrow \S \cdot \partial_{\mu} \S^{-1} = -(\partial_{\mu} b)\cdot \Z + \sin b \cdot \tilde{\B}_{\mu} +(1- \cos b) \cdot \P_{\mu} \; .
\end{equation}
Thus, in order to obtain the desired bundle reduction requirement (\ref{e5:38}), one fixes the group parameter $b$ as a constant over space-time ($\leadsto \partial_{\mu} b \equiv 0$) such that 
\begin{equation}
\label{e5:51}
\begin{array}
{l}\cos b = 0 \\ \sin b = 1\; ,
\end{array}
\end{equation}
and in this way one arrives at the desired result (for $\sigma_{\ast}=+1$) 
\begin{equation}
\label{e5:52}
\Bcirc_{\mu}=\P_{\mu}+\tilde{\B}_{\mu} \; .
\end{equation}
The $\GO(1,2)$-valued 1-form $\tilde{\B}_{\mu}$, emerging herein, is the desired complement $\E_{\mu}$ (\ref{e5:38}) of the density connection $\B_{\mu}$, i.e. more concretely
\begin{equation}
\label{e5:53}
\tilde{\B}_{\mu} \doteqdot [\B_{\mu},\Z]= \epsilon^l{}_{jk}B^j{}_{\mu}\theta^k \B_l \;,
\end{equation}
and thus it obeys the orthogonality condition in a trivial way
\begin{equation}
\label{e5:54}
tr(\Z \cdot \tilde{\B}_{\mu})=0 \; .
\end{equation}
Indeed, this is necessary in order that our present result $\Bcirc_{\mu}$ (\ref{e5:52}) has vanishing component in the holonomy algebra, see the former constraints (\ref{e5:47}) and (\ref{e5:32}).

In the last step, it remains to prove the surface relation (\ref{e5:48}) for the two-parted hyperboloid $H_{(+)}$; this however is easily achieved by simply comparing the general trivialization relation (\ref{e5:38}) for the change of connections with the present result (\ref{e5:52}), being due to our guess (\ref{e5:49}) - (\ref{e5:50}) for the positive mixtures. This comparison immediately yields namely for the algebraic complement ($\E_{\mu}$) of $\P_{\mu}$ in $\GO(1,2)$ 
\begin{equation}
\label{e5:55}
\E_{\mu} \equiv \tilde{\B}_{\mu} \; .
\end{equation}
And finally it is a nice exercise to substitute this back into the left-hand side of the surface relation (\ref{e5:48}) in order to be convinced of the validity of this formula. Simultaneously, the present method of bundle trivialization demonstrates that the Euler class $E_{\mu \nu}$ is an element of the {\it integral} cohomology group $H^2(M_4, {\mathbb{Z}})$, since the winding numbers $z[C^2]$ (\ref{e5:4}) must necessarily appear as integers. The reason for this is that the winding numbers may be normalized by means of the surface content of the (compact) range of the density map ($\thetavec$). (If the image space of $\thetavec$ is non-compact, as for the present situation of fig.1, nomalization of the winding numbers is not possible and they become all zero).

Thus the validity of the topological claim (\ref{e5:7}) has been proven for the positive mixtures ($\sigma_{\ast}=+1$) but not for the negative ones ($\sigma_{\ast}=-1$).

\subsection{Negative Mixtures}
But why does the present mechanism of bundle trivialization not work for the negative mixtures in an analogous way?

The reason is that the negative-mixture bundle ($\thetavec$) cannot be embedded into an $SO(1,2)$ bundle with trivial connection $\Bcirc_{\mu}$ (\ref{e5:38}). This is most easily seen by trying for the principal-bundle section $\Lambda(x)$ (\ref{e5:49}) the same $SO(1,2)$-valued ansatz $\S(x)$ (\ref{e5:20}) but with the density vector $\thetavec(x)$ being now ``space-like'', i.e. one takes $\sigma_{\ast}=-1$ for the normalization condition (\ref{e5:9}). This would namely yield for the corresponding pure gauge $\Bcirc_{\mu}$ (\ref{e5:49})
\begin{equation}
\label{e5:56}
\Bcirc_{\mu} \Rightarrow \S \cdot \partial_{\mu} \S^{-1} = -(\partial_{\mu} b)\cdot \Z + \sinh b \cdot \tilde{\B}_{\mu} +(1- \cosh b)\cdot \P_{\mu} \; .
\end{equation}
From here it is immediately seen that one has to take again the group parameter $b$ as a constant over space-time in order that the projection of $\Bcirc_{\mu}$ onto the holonomy algebra \{$\Z$\} be zero, cf. (\ref{e5:47}). However one evidently can have no non-trivial value for this constant $b$ so that the decomposition (\ref{e5:38}) does apply with both the connection forms $\Bcirc_{\mu}$ and $\B_{\mu}$ obeying the {\it inhomogeneous} law (\ref{e5:40}) and $\E_{\mu}$ obeying the {\it homogeneous} law (\ref{e5:39})! This is the true reason why the negative-mixture bundles cannot be embedded into a trivial $SO(1,2)$ bundle, as is the case with the positive-mixture configurations, and therefore we have to look for an even higher-dimensional trivial bundle for the purpose of embedding.

The solution to this problem is obtained by considering a trivial Lorentz bundle over space-time which has the proper Lorentz group $SO(1,3)$ as its structure group. The corresponding Lorentz algebra is easily obtained by simply complementing its $\GO(1,2)$ subalgebra, being spanned by the generators $\B_j$ (\ref{e5:10}) - (\ref{e5:12}), by three additional generators $\T_j$ which obey the following algebra
\begin{subequations}
\label{e5:57} 
\begin{align}
[\T_j,\T_k]&=-\epsilon^l{}_{jk}\B_l \\
[\T_j,\B_k]&=\epsilon^l{}_{jk}\T_l \; .
\end{align}
\end{subequations}
The Lorentz algebra $\GO(1,3)$ is usually not referred to the present $\GO(1,2)$ basis \{$\B_j$\} and its complement \{$\T_j$\} but rather to the ordinary $SO(3)$ rotation generators $L_j$ ($j=1,2,3$)
\begin{equation}
\label{e5:58}
[L_1,L_2]=L_3 \; \mbox{, (cycl.)}
\end{equation}
and the ``boost operators'' $l_j$:
\begin{subequations}
\label{e5:59} 
\begin{align}
[l_1,l_2]&=-L_3 \; \mbox{, (cycl.)}\\
[L_1,l_2]&=-[l_1,L_2]=l_3 \; \mbox{, (cycl.)}
\end{align}
\end{subequations}
(the remaining commutators of any subset are obtained by cyclic permutation (cycl.) of the generators). However it is easy to see the correspondence between both basis sets, namely:
\begin{equation}
\label{e5:60}
\begin{array}{ll}
\B_1 \Leftrightarrow -L_1 & \quad \quad\quad \quad \T_1 \Leftrightarrow l_1 \\ 
\B_2 \Leftrightarrow l_2 & \quad \quad \quad \quad \T_2 \Leftrightarrow L_2 \\
\B_3 \Leftrightarrow l_3 & \quad \quad \quad \quad \T_3 \Leftrightarrow L_3 \; .
\end{array}
\end{equation}
This embedding of the 3-dimensional $\GO (1,2)$ algebra into the 6-dimensional Lorentz algebra $\GO (1,3)$ leaves unchanged the bundle reduction formulae (\ref{e5:38})-(\ref{e5:48}) with the curvature being still given by (\ref{e5:36}) and the projector relation (\ref{e5:29}); and thus it is merely the definition of Hermiticity (\ref{e5:10}) which must be extended to the {\it four-dimensional} standard representation of the generators $\B_j=\{B^{\alpha}{}_{\beta j} \}$ and $\T_j=\{T^{\alpha}{}_{\beta j} \}$ ($\alpha ,\beta|=0,1,2,3$), i.e.
\begin{subequations}
\begin{align}
\label{e5:61}
\bar{B}^{\alpha}{}_{\beta j}&=\eta^{\alpha \gamma}B^{\delta}{}_{\gamma j} \eta_{\beta \delta}\\
\bar{T}^{\alpha}{}_{\beta j}&=\eta^{\alpha \gamma}T^{\delta}{}_{\gamma j} \eta_{\beta \delta}
\end{align}
\end{subequations}
\centerline{($\eta_{\alpha \beta}=diag(1,-1,-1,-1$)).}

According to this bundle embedding we try now again to find a trivial connection form $\Bcirc_{\mu}$ which takes its values in the embedding Lorentz algebra $\GO(1,3)$ and whose projection to the $\GO(1,2)$ subalgebra just yields the $\GO(1,2)$-valued connection $\B_{\mu}$ (\ref{e5:15}) for the {\it negative} mixture configurations ($\sigma_{\ast}=-1$). In other words, the desired connection $\Bcirc_{\mu}$ is taken again as the Maurer-Cartan form (\ref{e5:49}) but with the {\it global} section $\Lambda(x)$ now taking its values in the poper Lorentz group $SO(1,3)$:
\begin{equation}
\label{e5:62}
{\Lambda} \Rightarrow exp[b \theta^j \T_j].
\end{equation}
The associated Maurer-Cartan form $\Bcirc_{\mu}$ (\ref{e5:49}) is then easily computed as
\begin{equation}
\label{e5:63}
\Bcirc_{\mu} = -(\partial_{\mu}b)(\theta^j \T_j)+ \sin b \cdot \Bdt_{\mu}+ (1-\cos b)\cdot \P_{\mu}
\end{equation}
and thus is obviously the $SO(1,3)$ analogue of the former $SO(1,2)$ result (\ref{e5:50}) for the positive mixtures. The projective connection $\P_{\mu}$ is defined here again by the same prescription as for the lower-dimensional case (\ref{e5:26}) whereas the complement $\Bdt_{\mu}$, as the analogue of $\tilde{\B}_{\mu}$ (\ref{e5:53}) for the positive mixtures, is composed now of the complementary elements $\T_j$ alone:
\begin{equation}
\label{e5:64}
\Bdt_{\mu}=\epsilon^j{}_{k l} B^k{}_{\mu} \theta^l\T_j \; .
\end{equation}

Clearly, in order to get the connection $\B_{\mu}$ for the negative-mixure configurations as the $\GO (1,2)$ projection of the trivial $\Bcirc_{\mu}$ (\ref{e5:63}), one resorts again to the former arrangement (\ref{e5:51}) for the group parameter $b$, i.e. we have then the desired result for the negative mixtures
\begin{equation}
\label{e5:65}
\B_{\mu}=\Bcirc_{\mu} \Big|_{\GO (1,2)}
\end{equation}
and thus we obtain for the algebraic complement $\E_{\mu}$ (\ref{e5:38})
\begin{equation}
\label{e5:66}
\E_{\mu}=\Bdt \; .
\end{equation}
If this is substituted back into the general Euler class ${\bf{E}}$ (\ref{e5:45}) one arrives again at the original claim (\ref{e5:7}) whose validity for both kinds of mixtures is thus ensured, together with the topological conclusions being implied by that claim.

It is thus true that both kinds of mixtures share the triviality of their Euler class ${\bf{E}}$, yet there are nevertheless some geometric differences being worth to be considered in some more detail now.

\section{Discussion}

It is just the emergence of trivial principal bundles in connection with the 2-particle configurations which admits a concrete geometric picture of these
configurations. This appears in form of a local fibre distribution in the
tangent bundle. 

Let us consider the negative mixtures first, for which a {\em
  global} section $\Lambda(x)$ (\ref{e5:62}) exists for the embedding Lorentz
bundle. However since the latter bundle has the same fibre space $SO(1,3)$ as
the principal bundle associated with the tangent bundle of the underlying
space-time, one may look upon $\Lambda(x)$ also as a global section of the
(tangent) principal bundle, provided the existence of such global
sections is admitted by the topology of space-time. In order that this condition be satisfied, space-time must namely be a {\em 
  parallelisable} manifold (such as, e.g. the closed Robertson-Walker universe 
\cite{MiThoWhe} which is topologically the direct product of the 3-sphere $S^{3}$
and real line ${\mathbb{R}}^{1}$). By virtue of their very definition, parallelisable
manifolds admit the existence of a {\em global} frame \cite{GrKlMay,Hi}, i.e. for
the present case of four-dimensional space-time a tetrad of linearly
independent vector fields $\vec{f}_{\alpha}(x)\;(\alpha=0,1,2,3)$. These can
be chosen to be orthonormal with respect to the presumed space-time metric
${\bf g}$
\begin{equation}
\label{e6:1}
{\bf g}(\vec{f}_{\alpha},\vec{f}_{\beta})=\eta_{\alpha\beta}\;.
\end{equation}

The tetrad $\vec{f}_{\alpha}(x)$ may further be taken to be a solution of the transport equations
\begin{equation}
\label{e6:2}
\stackrel{\circ}{\nabla}_{\mu}\vec{f}_{\alpha}=\vec{f}_{\beta} \stackrel{\circ}{B^{\beta}}_{\alpha\mu}
\end{equation}
with the connection coefficients $\stackrel{\circ}{B^{\beta}}_{\alpha\mu}$ being due to the trivial connection $\Bcirc_{\mu}=\{\stackrel{\circ}{B^{\beta}}_{\alpha\mu}\}$ (\ref{e5:63}), which in general will be a non-metric connection. The vanishing of the curvature $\stackrel{\circ}{\mathcal{C}}_{\mu\nu}$ (\ref{e5:41}) of $\Bcirc_{\mu}$ then guarantees that any global tetrad solution $\vec{f}_{\alpha}(x)$ of (\ref{e6:2}) is unique over the whole space-time and thus it establishes a global isomorphism of all the tangent fibres. The restriction of the trivial Lorentz connection $\Bcirc_{\mu}$ to the non-trivial $\GO(1,2)$ connection $\B_{\mu}$ (\ref{e5:65}) geometrically means to omit one of the (space-like) tetrad vectors and to consider the remaining triad $\vec{f}_{j}(x)\;(j=0,1,2)$
\begin{equation}
\label{e6:3}
\nabla_{\mu}\vec{f}_{j}=\vec{f}_{k}B^{k}_{j\mu}\;.
\end{equation}
This triad defines a local (1+3) splitting of the tangent fibres. Since
however the curvature $\mathcal{C}_{\mu\nu}$ (\ref{e5:36}) of $\B_{\mu}$ is
proportional to the $\GO(1,1)$ generator $\Z$, it
annihilates the same vectors as does the projector $\P$ (\ref{e5:29}), which is nothing else than the square of the generator $\Z$. But this two-dimensional projector $\P$ itself
annihilates a 2-dimensional subspace of any tangent fibre and thus induces a local (2+2)-splitting. In the corresponding 2-distribution, as a sub-distribution of the former ($1+3$) splitting (\ref{e6:3}), the projector $\P$ acts as the identity and if this 2-distribution is
integrable, there arises a foliation of space-time into a system of 2-surfaces. The Euler class ${\bf E}$ (\ref{e5:1}) appears then as the intrinsic curvature
of these 2-surfaces. In this sense, the negative two-particle mixtures are
geometrically represented by a system of 2-surfaces which are equipped with a Lorentzian tangent metric. The existence of such a metric is guaranteed by the vanishing of the Euler number $z$ (\ref{e5:4}) \cite{NaSe}; and its invariance group consists of the $SO(1,1)$ tranformations being specified by equation (\ref{e5:33}).

On the other hand, the positive mixtures are geometrically characterized by a trivial $\GO(1,2)$ connection $\Bcirc_{\mu}$ (\ref{e5:50}) and thus start
with a global triad of tangent vectors $\vec{f}_{j}(x)$ from the very
beginning
\begin{equation}
\nabla_{\mu}\vec{f}_{j}=\vec{f}_{k}\stackrel{\circ}{B^{k}}_{j\mu}\;.
\end{equation}
However such a triad can always be complemented to a tetrad (by selection of
some linearly independent fourth vector) and then the same arguments do apply
as for the negative mixtures. The geometric difference between both mixtures
is that the representative 2-surfaces for the positive mixtures are space-like 
(such that the gauge element $\S$ (\ref{e5:33}) acts as an ordinary rotation of the
space-like tangent plane of the 2-surfaces) whereas the negative mixtures have 
a (1+1)-dimensional Minkowski plane as their tangent space.

\begin{figure}
\epsfig{file=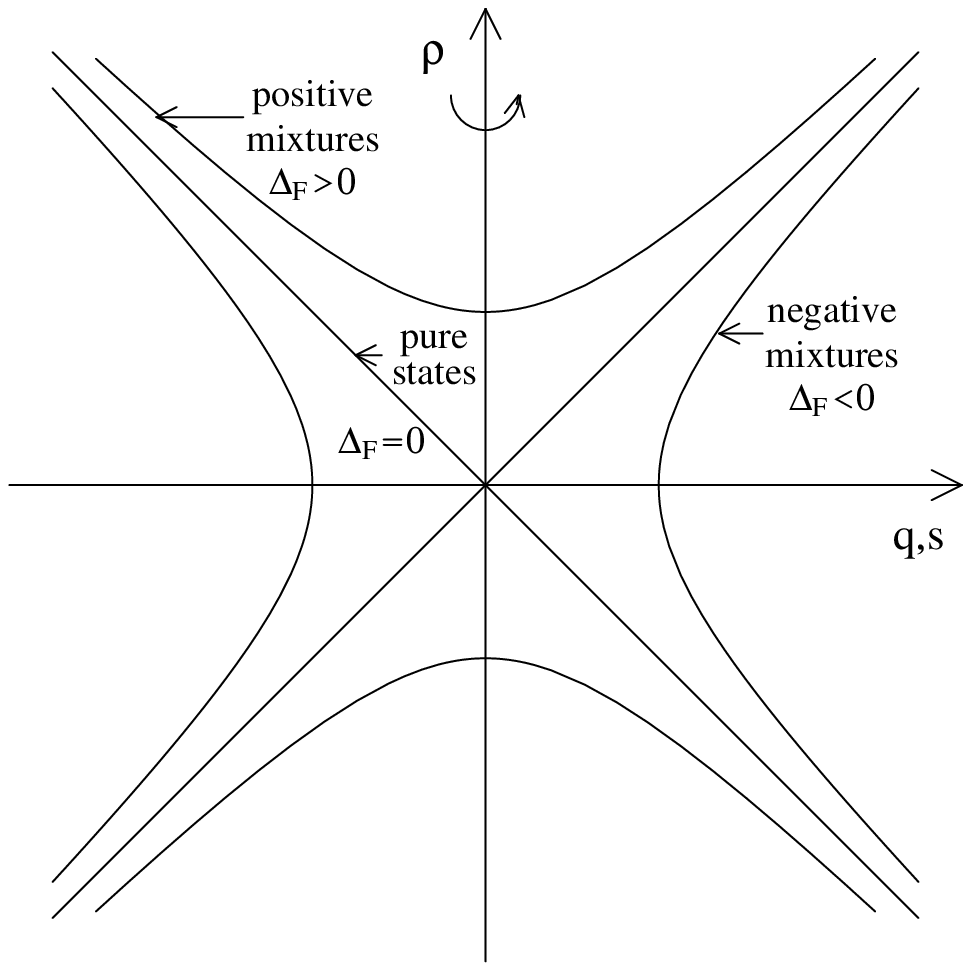}
\caption{The relativistic von Neumann equation (\ref{eq:calI}) subdivides the density configuration
space into three subsets: the {\em pure states} occupy the {\em Fierz
  cone} $(\Delta_{F}=0)$, {\em positive mixtures} $(\Delta_{F}>0)$ are geometrically
represented by the two-parted hyperboloid and the {\em negative mixtures}
$(\Delta_{F}<0)$ by the one-parted hyperboloid. The mixtures approach the
pure states for $\Delta_{F}\rightarrow 0$. The general RST dynamics forbids a
change of the mixture type, cf. (\ref{e4:19}). The positive (negative)
mixtures may be considered as the RST counterparts of the symmetric 
(anti-symmetric) states of the conventional quantum theory.
}
\label{fig1}    
\end{figure}

\end{document}